\documentclass[10pt, twocolumn]{article}

\usepackage{thermodynamic-clean}  
\usepackage{stfloats}
\usepackage{cleveref}
\crefname{section}{Section}{Sections}
\Crefname{section}{Section}{Sections}
\crefname{subsection}{Section}{Sections}
\Crefname{subsection}{Section}{Sections}
\crefname{subsubsection}{Section}{Sections}
\Crefname{subsubsection}{Section}{Sections}
\crefname{figure}{Fig.}{Figs.}
\Crefname{figure}{Fig.}{Figs.}
\crefname{table}{Table}{Tables}
\Crefname{table}{Table}{Tables}
\crefname{equation}{Eq.}{Eqs.}
\Crefname{equation}{Eq.}{Eqs.}
\crefname{appendix}{Appendix}{Appendices}
\Crefname{appendix}{Appendix}{Appendices}
\crefname{subappendix}{Appendix}{Appendices}
\Crefname{subappendix}{Appendix}{Appendices}

\renewcommand{\shorttitle}{CN101 - A Digital Thermodynamic Computer for Generative AI}

\begin{document}

\twocolumn[{%
\begin{@twocolumnfalse}
  \vspace*{4mm}
  {\LARGE\bfseries {\color{normalred}CN101} - A Digital Thermodynamic Computer for Generative AI}\\[4pt]
  {\large\itshape Targeting the Sequential Bottleneck of Modern Generative AI}\\[10pt]
  {\normalsize
  Lars Holdijk$^{1, 2}$,
  Denis Melanson$^{1}$,
  Zier Mensch$^{3,4}$,
  Brandon Birchall$^{1}$,
  Vincent Cheung$^{1}$,
  Nicholas Lehrter$^{1}$,
  Maxwell Aifer$^{1}$, \\
  Samuel Duffield$^{1}$,
  Jan Ole Ernst$^{1}$,
  Rajath Salegame$^{1}$,
  Antonio J. Martinez$^{1}$,
  Gavin Crooks$^{1}$,
  Miranda Cheng$^{3,5}$,
  Zach Belateche$^{1}$,
  Marc~Bright$^{1}$,
  Patrick J. Coles$^{1}$,
  Faris Sbahi$^{1}$
  }\\[4pt]
  {\small
    $^{1}$ Normal Computing Corporation, New York, USA\\
    $^{2}$ University of Oxford, Oxford, United Kingdom\\
    $^{3}$ University of Amsterdam, Amsterdam, Netherlands\\
    $^{4}$ National Taiwan University, Taipei, Taiwan\\
    $^{5}$ Academia Sinica, Taipei, Taiwan\\
    Correspondence:
    \href{mailto:research@normalcomputing.com}{research@normalcomputing.com},  \href{mailto:larsholdijk@gmail.com}{larsholdijk@gmail.com}
    \vspace{-.5cm}
  }\\[4pt]
  \brandingrule
  \vspace{2pt}
\begin{abstractbox}
\textbf{\color{normalred}Abstract}\quad
Thermodynamic computing is an emerging hardware paradigm, in which stochastic physical dynamics serve as the direct computational primitive. The recent explosion of generative AI has only sharpened the search for alternative approaches to compute, and, as we show in this work, thermodynamic computing turns out to be well suited to this space. An important class of methods realises a function $f(x)$ as the stationary expectation of an ergodic stochastic process: the answer is encoded in the time-averaged statistics of an equilibrating trajectory. To date, this equilibration-style class has been formulated exclusively through Langevin dynamics, restricting its implementations to analogue substrates and the engineering challenges those bring. In this work, we propose a substrate-independent formalisation of the equilibration-style formulation, in which the only object of design is the dynamical generator $\Lstar$ of an arbitrary ergodic process. The formalisation makes three hardware-level properties of the formulation explicit: the precision of a result is a knob set by how long the dynamics are run, sample averages decompose across independent trajectories, and dependent stages of a computation operate concurrently rather than serially, a property we call sequential parallelism. We instantiate the formalisation by fabricating a prototype digital thermodynamic computing chip, named CN101, that implements the formulation through discrete accumulator dynamics on standard CMOS using stochastic computing principles. We characterise CN101's success across conventional generative AI workloads in the form of VAEs and flow matching, applied to both image generation and scientific problems.
Together, the formalisation and its digital instantiation show that the equilibration-style formulation is substrate-independent, and that its computational properties can be exploited on standard digital hardware.
\end{abstractbox}

  \vspace{4mm}
\end{@twocolumnfalse}
}]

\section{Introduction}
In recent years, there has been growing interest in thermodynamic computing as an alternative hardware paradigm\cite{conte2019,coles2023thermodynamic_published,hylton2020,aifer2025solving}. The explosion of generative AI has only sharpened this interest: as these workloads strain conventional accelerators, the search for alternative approaches to compute has intensified, and thermodynamic computing turns out to be well suited to them. The unifying premise across the field is that stochastic physical dynamics can serve as a direct computational primitive \cite{conte2019}. Rather than fighting noise to implement deterministic logic, thermodynamic computing uses stochasticity as part of the computational process. Within this broad concept, distinct computational styles have developed, each defining its own relationship between physical dynamics and computational output.

\begin{figure}[t!]
    \vspace{-.2em}
    \centering
    \includegraphics[width=\linewidth]{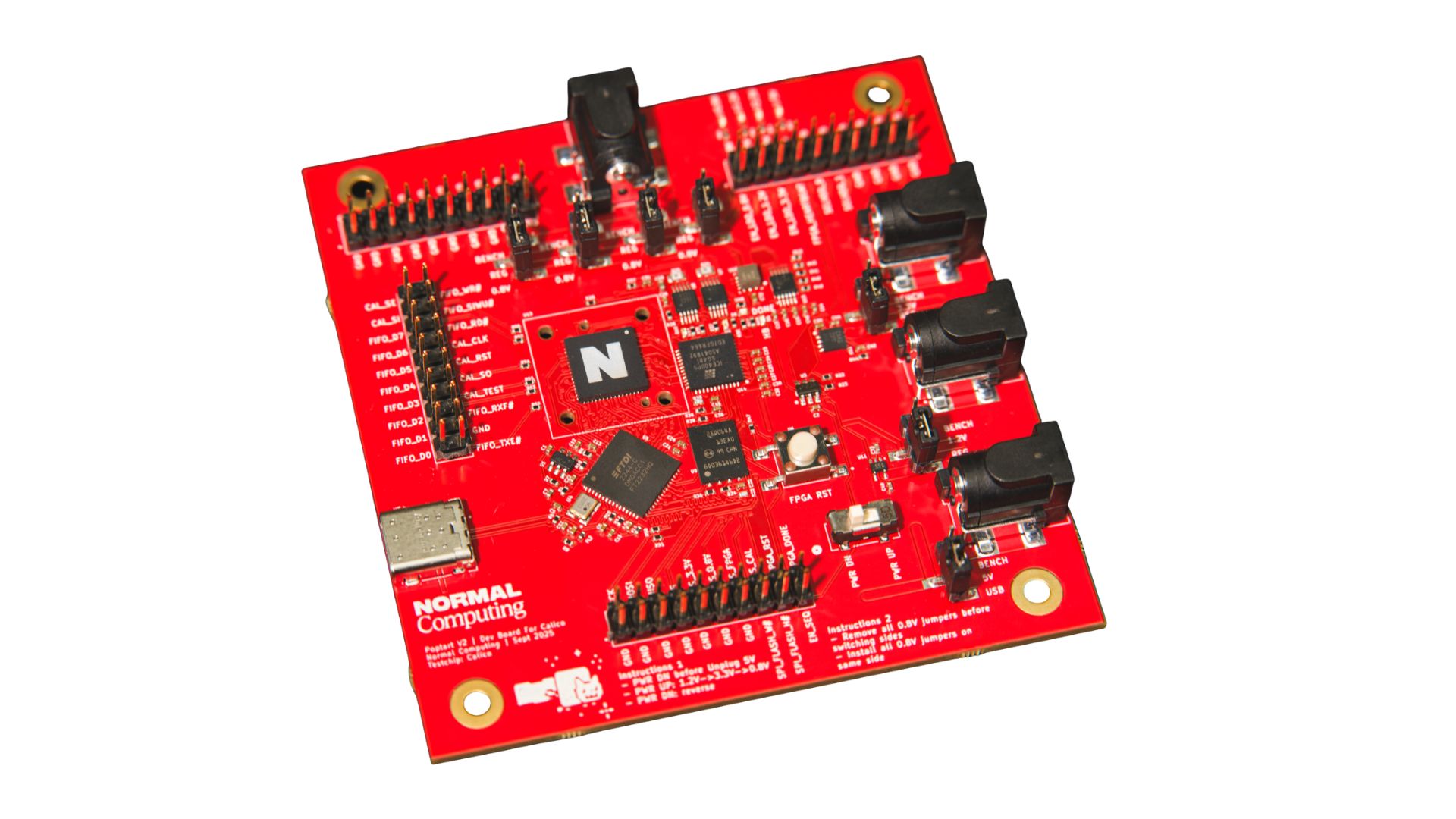}
  \caption{The evaluation PCB carries CN101 together with an FPGA, power regulation, and a USB-C host interface.}
    \label{fig:chip:pcb}
    \vspace{-.2em}
\end{figure}
In this work, we focus on the class of methods that realises a function $f(x)$ as the stationary expectation of an ergodic stochastic process\cite{aifer2024_TLA,melanson2025}, an equilibration-style formulation in which the time-averaged statistics of a stochastic trajectory encode the answer. This class was established by Thermodynamic Linear Algebra (TLA)\cite{aifer2024_TLA,melanson2025} and validated in analogue hardware by the Stochastic Processing Unit (SPU)\cite{melanson2025}; we describe the underlying dynamics in \cref{sec:background}. The same formulation has also been carried to further problem classes, such as Bayesian inference\cite{aifer2024thermodynamicbayes} and natural gradient descent\cite{donatella2026thermodynamicngd}. Subsequent work has extended the formulation to nonlinear and non-equilibrium
settings\cite{whitelam2026nonlinear,whitelam2026generative}, including generative modelling by reverse-time evolution of analogue Langevin systems\cite{whitelam2026generative}. Distinct from this line of work, a separate direction of thermodynamic computing has developed around discrete probabilistic bits sampling energy-based models\cite{camsari2017pbit,kaiser2021pbit,aadit2022,coles2023thermodynamic_published,extropic2025}. Methods here are defined by a different computational formulation where the focus is on sampling from a programmable probability distribution rather than evaluating arbitrary functions using a thermodynamic process, as is done in the equilibration-style methods that motivate this work. We do not address the sampling focused direction here.

Despite the rapid advances over the last years, a unifying definition for the equilibration-style formulation has not been written down independent of its analogue Langevin instances that have motivated the thermodynamic hardware development so far. Instead, each demonstration of the field carries its own implicit computational contract, with the consequence that each formulation has remained tied to a specific physical implementation. The most immediate practical consequence is that thermodynamic hardware has primarily been confined to analogue substrates, which bring well-known engineering challenges, such as device variability, limited dynamic range, and a manufacturing path that does not benefit from the established CMOS toolchain.

To address this, we make three contributions. First, we give a substrate-independent formal definition of equilibration-based computation, in which the dynamical generator of an ergodic process is the only object of design. The analogue Langevin systems of prior work appear as one special case here (\cref{sec:framework}), but the definition also allows for other formats such as discrete Markov chains. Second, we characterise three scaling properties of thermodynamic computation (anytime precision, parallel sample aggregation, and sequential parallelism) that are exploitable in hardware independently of the substrate. And third, to instantiate this formalisation in silicon, we present a digital thermodynamic computing chip, codenamed CN101, based on stochastic computing principles~\cite{gaines1969,alaghi2013}.

CN101 implements the dynamical generator $\Lstar$ at the core of the presented thermodynamic computing formulation through accumulator-based finite-state machines updated by pseudo-random bit streams. As the chip is built on standard CMOS, its design directly benefits from established manufacturing and design tooling, in contrast to the analogue implementations of prior work. Notably, CN101 is a prototype built to prove out the substrate-independent formulation in silicon rather than as a finished or commercial design, and many of its blocks are deliberately first-generation, with substantial improvements left to subsequent chips in the Carnot Architecture. \cref{fig:chip:pcb} shows CN101 mounted on its evaluation PCB.

To validate CN101's operation we perform extensive testing across multiple workloads. First, we confirm that CN101 reproduces the formulation's original problem class, solving linear systems, in fully digital form. We then turn to a conditional variational autoencoder generating MNIST digits, a nonlinear generative model run entirely on-chip and use it to characterise the discussed scaling properties of thermodynamic computing. Using multiple chips, we then shift our focus to modern generative AI methods and show CN101 operating on problems within image generation and the life sciences. Every model we deploy is a standard architecture of the kind trained and run on GPUs (a conditional variational autoencoder and flow-matching models) without specialised modifications.



\section{Substrate Independent Thermodynamic Computing}
\label{sec:framework}


In the coming section we will develop the substrate independent formulation of thermodynamic computing, discuss the core challenges in implementing this framework, and outline three key characteristics and the benefits they bring when successful. First, however, we quickly discuss the fundamental direction in this branch of thermodynamic computing built around equilibration.

\subsection{Background: Equilibrium Thermodynamic Computing}\label{sec:background}
Thermodynamic computing takes its name from a structural analogy with statistical mechanics. A physical system in contact with a heat bath relaxes to thermal equilibrium, where the Boltzmann distribution $\pi(S) \propto e^{-H(S)/k_B T}$ assigns a probability to each configuration $S$ and macroscopic observables are equilibrium expectations $\mathbb{E}_\pi[\varphi(S)]$ of microscopic quantities $\varphi$. These averages are never evaluated by an explicit sum over configurations; they \emph{emerge}, in the sense that once the system has equilibrated, a time average of $\varphi$ along a single trajectory converges to the ensemble expectation. Thermodynamic computing inverts this relationship: rather than reading out the equilibrium properties of a given system, one engineers a system whose equilibrium expectation is a quantity one wishes to compute.

A clear realisation of this idea is Thermodynamic Linear Algebra (TLA)~\cite{aifer2024_TLA,melanson2025}. To solve a linear system $As = b$ with symmetric positive-definite $A$, one runs the Ornstein--Uhlenbeck dynamics~\cite{uhlenbeck1930}
\begin{equation}
  \mathrm{d}S_t = -(A S_t - b)\,\mathrm{d}t + \sqrt{2}\,\mathrm{d}W_t,
  \label{eq:ou_tla}
\end{equation}
whose stationary distribution is the Gaussian $\mathcal{N}(A^{-1}b,\, A^{-1})$. Its two moments give two linear-algebra primitives: the mean $A^{-1}b$ is the solution of the system, recovered as the time-averaged state $\tfrac{1}{T}\int_0^T S_t\,\mathrm{d}t$, and the covariance $A^{-1}$ is the matrix inverse, recovered from the sample covariance of the state. Neither requires a matrix factorisation. The Stochastic Processing Unit (SPU)~\cite{melanson2025} realised these dynamics in analogue hardware as a set of capacitively coupled RLC cells, with $A$ encoded in the cells' tunable capacitances and the cells driven by an injected current-noise source sampling the equilibrium voltages and forming their covariance; it inverted dense matrices in hardware.

Two features of this construction are worth noting, as both carry over to the substrate-independent setting developed below. First, the computation is performed by relaxation rather than by an explicit algorithm: the SPU inverts a $d \times d$ matrix in a time that scales as $\mathcal{O}(d^2)$, against the $\mathcal{O}(d^3)$ of direct factorisation~\cite{aifer2024_TLA}. Second, the readout is available at any time. The running average is already a usable estimate at every $T$ and improves as the process continues. We develop this property into a general scaling axis in \cref{sec:scaling}. What has so far been specific to TLA is the choice of Langevin dynamics on a continuous space, and it is precisely this choice that ties these realisations to analogue hardware. The formalism developed removes this restriction.

\subsection{The Equilibration Formalism}\label{sec:formalism}
We now state this pattern as a definition that makes no commitment to any particular dynamics or physical substrate:

\begin{contractbox}
A thermodynamic computation of $f : \RR^n \to \RR^m$ is specified by a tuple $(S, \Lstar, \pi, \varphi)$, where $S$ is a state space, $\Lstar$ is a dynamical generator parameterised by $f$ and the input $x$, $\pi$ is the unique stationary distribution satisfying $\Lstar\pi = 0$, and $\varphi : S \to \RR^m$ is a readout observable with $\Epi[\|\varphi(S)\|]<\infty$. The computation is
\begin{equation}
  f(x) \;\triangleq\; \Epi[\varphi(S)],
  \label{eq:contract}
\end{equation}
and the output is the time-averaged readout
\begin{equation}
  \hat{y}_T = \frac{1}{T}\sum_{t=1}^T \varphi(S_t),
  \label{eq:time_average}
\end{equation}
which converges almost surely to $f(x)$ as $T \to \infty$ whenever $\Lstar$ generates ergodic dynamics~\cite{meyn2009markov}, regardless of initial state.
\end{contractbox}

To make the tuple concrete, take the linear system $Ax = b$ of \cref{sec:background}. Here the state space $S$ is that of the Ornstein--Uhlenbeck process, the generator $\Lstar$ is the operator of the dynamics $\mathrm{d}S_t = -(A S_t - b)\,\mathrm{d}t + \sqrt{2}\,\mathrm{d}W_t$ with $A$ and $b$ encoding the input, and the stationary distribution is the Gaussian $\pi = \mathcal{N}(A^{-1}b,\,A^{-1})$. Taking the identity readout $\varphi(S) = S$ gives $\Epi[\varphi(S)] = A^{-1}b$, so the time-average $\hat{y}_T$ of the state is the running estimate of the solution, and reading the sample covariance instead recovers the inverse $A^{-1}$.

What the definition requires is only that the dynamics equilibrate to a unique stationary distribution. The vocabulary of energies and temperatures motivates the name and supplies the intuition, but it plays no role in the definition itself, and the generator need not correspond to any physical system.

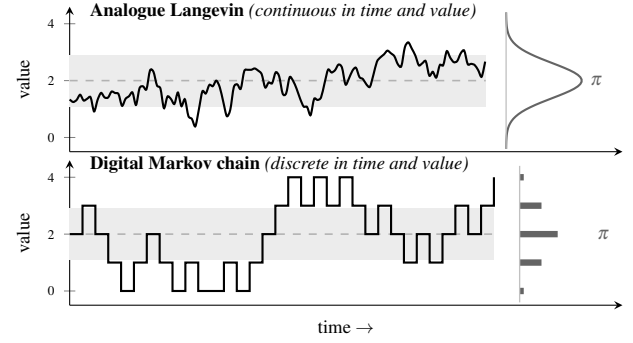
\begin{figure}[t]
\centering
\begin{tikzpicture}
\begin{axis}[
  name=top, width=\columnwidth, height=3.5cm,
  xmin=0, xmax=19, ymin=-0.5, ymax=4.6,
  xtick=\empty, ytick={0,2,4}, ytick align=outside,
  axis lines=left, tick label style={font=\tiny}, label style={font=\scriptsize},
  ylabel={value}, xlabel={}, clip=false,
  title style={font=\scriptsize\bfseries, at={(0.02,0.98)}, anchor=north west},
  title={Analogue Langevin \textmd{\itshape(continuous in time and value)}},
]
  \fill[normalgray!12] (axis cs:0,1.1) rectangle (axis cs:14.3,2.9);
  \addplot[draw=normalgray!50, dashed, line width=0.6pt] coordinates {(0,2)(14.3,2)};
  \addplot[normalred, line width=0.8pt, smooth] coordinates {(0.0,1.336) (0.12,1.247) (0.24,1.32) (0.36,1.504) (0.48,1.3) (0.6,1.36) (0.72,1.413) (0.84,0.915) (0.96,1.324) (1.08,1.568) (1.2,1.409) (1.32,1.405) (1.44,1.613) (1.56,1.564) (1.68,1.525) (1.8,1.112) (1.92,1.359) (2.04,1.452) (2.16,1.583) (2.28,1.142) (2.4,1.729) (2.52,1.8) (2.64,1.696) (2.76,2.354) (2.88,2.31) (3.0,1.832) (3.12,1.72) (3.24,1.03) (3.36,1.438) (3.48,1.356) (3.6,1.178) (3.72,1.582) (3.84,1.102) (3.96,1.344) (4.08,0.756) (4.2,0.654) (4.32,0.392) (4.44,0.983) (4.56,1.619) (4.68,1.548) (4.8,1.848) (4.92,1.805) (5.04,1.998) (5.16,1.764) (5.28,1.251) (5.4,0.752) (5.52,0.976) (5.64,1.763) (5.76,1.867) (5.88,1.714) (6.0,2.334) (6.12,2.38) (6.24,2.38) (6.36,2.427) (6.48,2.35) (6.6,2.224) (6.72,1.758) (6.84,1.934) (6.96,1.91) (7.08,2.29) (7.2,2.151) (7.32,1.544) (7.44,1.551) (7.56,2.118) (7.68,1.989) (7.8,1.712) (7.92,1.364) (8.04,1.09) (8.16,1.073) (8.28,0.783) (8.4,1.352) (8.52,1.319) (8.64,1.41) (8.76,1.908) (8.88,2.384) (9.0,2.286) (9.12,2.365) (9.24,2.564) (9.36,2.456) (9.48,1.863) (9.6,2.079) (9.72,2.351) (9.84,2.451) (9.96,2.125) (10.08,2.054) (10.2,1.865) (10.32,1.783) (10.44,2.206) (10.56,2.665) (10.68,2.818) (10.8,2.92) (10.92,3.054) (11.04,2.962) (11.16,2.857) (11.28,2.575) (11.4,2.51) (11.52,3.171) (11.64,3.344) (11.76,3.124) (11.88,2.883) (12.0,2.539) (12.12,2.61) (12.24,2.681) (12.36,2.174) (12.48,2.311) (12.6,2.107) (12.72,2.543) (12.84,2.546) (12.96,3.036) (13.08,2.806) (13.2,2.649) (13.32,2.688) (13.44,2.959) (13.56,3.055) (13.68,2.582) (13.8,2.59) (13.92,2.547) (14.04,2.368) (14.16,2.135) (14.28,2.668)};
  \addplot[normalgray, line width=0.8pt] coordinates {(15.001,-0.4) (15.002,-0.319) (15.004,-0.237) (15.006,-0.156) (15.01,-0.075) (15.015,0.007) (15.022,0.088) (15.033,0.169) (15.049,0.251) (15.07,0.332) (15.098,0.414) (15.137,0.495) (15.186,0.576) (15.249,0.658) (15.329,0.739) (15.425,0.82) (15.541,0.902) (15.677,0.983) (15.833,1.064) (16.006,1.146) (16.195,1.227) (16.396,1.308) (16.602,1.39) (16.807,1.471) (17.004,1.553) (17.184,1.634) (17.34,1.715) (17.464,1.797) (17.55,1.878) (17.594,1.959) (17.594,2.041) (17.55,2.122) (17.464,2.203) (17.34,2.285) (17.184,2.366) (17.004,2.447) (16.807,2.529) (16.602,2.61) (16.396,2.692) (16.195,2.773) (16.006,2.854) (15.833,2.936) (15.677,3.017) (15.541,3.098) (15.425,3.18) (15.329,3.261) (15.249,3.342) (15.186,3.424) (15.137,3.505) (15.098,3.586) (15.07,3.668) (15.049,3.749) (15.033,3.831) (15.022,3.912) (15.015,3.993) (15.01,4.075) (15.006,4.156) (15.004,4.237) (15.002,4.319) (15.001,4.4)};
  \draw[normalgray!40,line width=0.5pt] (axis cs:15,-0.4)--(axis cs:15,4.4);
  \node[normalgray, font=\scriptsize] at (axis cs:18.1,2){$\pi$};
\end{axis}
\begin{axis}[
  name=bot, at={(top.below south west)}, anchor=north west, yshift=-3pt,
  width=\columnwidth, height=3.5cm,
  xmin=0, xmax=43, ymin=-0.5, ymax=4.6,
  xtick=\empty, ytick={0,2,4}, ytick align=outside,
  axis lines=left, tick label style={font=\tiny}, label style={font=\scriptsize},
  ylabel={value}, xlabel={time $\rightarrow$}, clip=false,
  title style={font=\scriptsize\bfseries, at={(0.02,0.98)}, anchor=north west},
  title={Digital Markov chain \textmd{\itshape(discrete in time and value)}},
]
  \fill[normalgray!12] (axis cs:0,1.1) rectangle (axis cs:33,2.9);
  \addplot[draw=normalgray!50, dashed, line width=0.6pt] coordinates {(0,2)(33,2)};
  \addplot[normalred, line width=0.8pt, const plot mark left] coordinates {(0,2) (1,3) (2,2) (3,1) (4,0) (5,1) (6,2) (7,1) (8,0) (9,1) (10,0) (11,0) (12,1) (13,0) (14,1) (15,2) (16,3) (17,4) (18,3) (19,4) (20,3) (21,4) (22,3) (23,2) (24,3) (25,2) (26,1) (27,2) (28,1) (29,2) (30,3) (31,2) (32,3) (33,4)};
  \draw[normalgray,line width=2.4pt] (axis cs:35,0) -- (axis cs:35.32,0);
      \draw[normalgray,line width=2.4pt] (axis cs:35,1) -- (axis cs:36.7,1);
      \draw[normalgray,line width=2.4pt] (axis cs:35,2) -- (axis cs:37.96,2);
      \draw[normalgray,line width=2.4pt] (axis cs:35,3) -- (axis cs:36.7,3);
      \draw[normalgray,line width=2.4pt] (axis cs:35,4) -- (axis cs:35.32,4);
        \draw[normalgray!40,line width=0.5pt] (axis cs:35,-0.4)--(axis cs:35,4.4);
  \node[normalgray, font=\scriptsize] at (axis cs:41.5,2){$\pi$};
\end{axis}
\end{tikzpicture}
\vspace{-.4em}
\caption{\textbf{Different forms of Equilibrium Thermodynamic Computing.} Different stochastic processes can relax to the same $\pi$, and so compute the same $f(x)=\Epi[\varphi(S)]$ under the equilibrium definition of thermodynamic computing. \textbf{Top:} an analogue Langevin process wanders continuously, in both time and value. \textbf{Bottom:} a digital Markov chain takes discrete steps between discrete states.}
\vspace{-.9em}
\label{fig:framework}
\end{figure}

This is what makes the definition substrate-independent. The realisation of \cref{sec:background} took $\Lstar$ to be a Langevin operator on a continuous state space, which requires continuous noise and so confines them to analogue hardware~\cite{melanson2025}. The definition given here, however, does not impose such a restriction: $\Lstar$ may be any ergodic generator of which the stationary distribution encodes the computation, including a discrete-state Markov chain realised entirely in digital logic (\cref{fig:framework}).

\subsubsection{Challenges}\label{sec:challenges}
Realising this computation for a given $f$ in a hardware-native approach requires solving three coupled challenges related to the encoding, implementation, and efficiency.

\paragraph{Encoding: }The encoding problem is to find a generator $\Lstar$ and readout $\varphi$ such that $\Epi[\varphi(S)] = f(x)$, demanding co-design of all three objects: generator, observable, and input encoding. For TLA this is known analytically: an Ornstein--Uhlenbeck process with $A$ encoded in the drift has a Gaussian stationary distribution with mean $A^{-1}b$, and the identity readout suffices~\cite{aifer2024_TLA}. For general $f$, such as the input-output map of a neural network, no trivial recipe exists.

\paragraph{Implementation: }Beyond encoding, the generator must also be implementable as the native dynamics of a physical or digital substrate, not as a simulation running on top of a general-purpose processor. The SPU is the canonical example: its RLC circuit does not simulate the Ornstein--Uhlenbeck process but is that process, with the dynamics carried out by the device's physics rather than by instructions on a host. This is what distinguishes thermodynamic computing from running Markov chain Monte Carlo on a GPU: on a GPU the substrate bears no structural relationship to the generator, whereas in thermodynamic computing the structure of the substrate \emph{is} the structure of the generator.

\paragraph{Efficiency: }Even with the correct stationary distribution and a realisable generator, the dynamics must also reach and then explore that distribution fast enough for $\hat{y}_T$ to converge in useful time. This is governed by two principles, burn-in and mixing time, as familiar from Markov chain Monte Carlo~\cite{brooks2011handbook}.

The first is \emph{burn-in}: the chain must reach its stationary distribution at all, and the number of steps it takes to approach $\pi$ is the relaxation time $\trelax$. Below $\trelax$ the readout carries a systematic bias from the initial state that decays only as $T$ grows past $\trelax$. The second is \emph{mixing}: once at equilibrium, successive samples along a trajectory must decorrelate for average to improve and reduce variance. The rate of this is set by the correlation time $\tcorr$. A hardware implementation that encodes the right distribution but either equilibrates or mixes slowly, with $\trelax$ or $\tcorr$ large, is uncompetitive.

\subsection{Scaling Axes of Thermodynamic Computing}\label{sec:scaling}
Because every output is a statistical average over a stochastic trajectory rather than the result of a fixed instruction sequence, an equilibration-based computation can be scaled along axes that have no clean analogue on conventional accelerators. We describe three. The first trades time for accuracy within a single run, the second trades hardware for accuracy across many runs, and the third removes the wall-clock cost of composing dependent computations. All three follow directly from the statistical nature of the readout, and each maps onto a concrete lever for scheduling work on real hardware.

\subsubsection{Anytime precision.}
The readout $\hat{y}_T$ is an average over $T$ samples, and once past the burn-in for $T$ beyond $\trelax$, its error falls as the $\mathcal{O}(T^{-1/2})$ of the mixing regime~\cite{meyn2009markov}. Precision is therefore not a fixed property of the hardware but a knob set by how long the process is run past the initial burn-in: the computation returns a valid, if coarse, answer at every $T$ and refines it monotonically in expectation as $T$ grows~\cite{zilberstein1996}. A workload can be run only until it reaches the accuracy the downstream task requires and then terminated, rather than to some hardware-fixed bit width, so the energy that would have bought further precision is simply never spent.

In a generative setting the same lever permits early exit: a draft can be inspected as it sharpens and abandoned before it is fully formed if it is not worth completing. Unlike a fixed-latency accelerator, which must finish all steps in computation before producing any output, an equilibration-based computation makes precision, time, and energy a tunable trade-off.

\subsubsection{Parallel sample aggregation.}
A single run carries an intrinsic cost in the variance of the readout from one replica set by the variance of the observable under $\pi$ and the correlation time $\tcorr$ of the generator,  $\mathrm{Var}(\hat{y}_T) \approx \tcorr\,\mathrm{Var}_\pi(\varphi)\,/\,T$~\cite{meyn2009markov}. How quickly this base variance falls is precisely the efficiency challenge of \cref{sec:challenges}. Fortunately, parallel sample aggregation provides a lever here, allowing one to trade off variance with extra hardware commitment.

This lever comes from the observation that trajectories produced by independent runs of the same dynamics are statistically independent, so $K$ replicas of the hardware running in parallel produce a combined estimator whose variance is $K$ times smaller than a single replica of the same wall-clock duration, with no inter-replica coordination beyond a final averaging step~\cite{brooks2011handbook}.

\begin{contractbox}
The aggregate estimator
\begin{equation}
  \bar{y}_T^{(K)} = \frac{1}{K}\sum_{k=1}^K \hat{y}_T^{(k)}
  \label{eq:parallel_agg}
\end{equation}
satisfies $\mathrm{Var}(\bar{y}_T^{(K)}) = \mathrm{Var}(\hat{y}_T^{(1)})/K$, so the rate of variance reduction scales with the aggregate throughput $KT$ rather than either factor alone.
\end{contractbox}

This makes the precision of a computation a matter of resource allocation. In a datacentre, more chips or more power can be directed at a single latency-critical computation to reach a target accuracy sooner, and withdrawn again when that additional computation is no longer needed, with the variance of the result tracking the resources assigned to it.

It needs to be noted here that there is a limit to what more hardware buys. Aggregation drives down the variance but leaves the burn-in bias untouched. At a fixed per-replica window $T$ this can be averaged down only as far as the systematic floor that the burn-in leaves behind. Past that floor, precision is recovered not by adding replicas but by rolling each of them out beyond $\trelax$.

\subsubsection{Sequential parallelism.}
Of the three axes, this is perhaps the most consequential for generative AI, where the workloads of interest are long chains of dependent stages. Suppose the target function is realised not as a single thermodynamic computation but as a composition $f = f_L \circ \cdots \circ f_1$ of $L$ stages, each $f_\ell$ itself a thermodynamic computation in the sense of \cref{sec:formalism}, with its own running estimate $\hat{y}_T^{(\ell)}$.

Each such estimate is available at every $T$ and is a consistent estimator of its stage's stationary expectation, as described by the anytime precision property. As such, a downstream stage $f_\ell$ can consume the running estimate $\hat{y}_T^{(\ell-1)}$ of the stage feeding it before that estimate has converged: early estimates are imprecise and improve over time, and as long as $f_\ell$ is continuous in its input, the downstream stage's output tracks the converging upstream estimate and remains consistent with the composed function $f$.

\begin{contractbox}
The continuous-mapping theorem~\cite{vanderVaart1998} states that if $X_n \to X$ almost surely and $g$ is continuous at $X$, then $g(X_n) \to g(X)$ almost surely. Applied along an $L$-stage pipeline in which each stage stays consistent as the estimate feeding it converges, it gives
\begin{equation}
  \hat{y}_T^{(L)} \;\xrightarrow{\mathrm{a.s.}}\; (f_L \circ \cdots \circ f_1)(x) \quad \text{as } T \to \infty,
  \label{eq:seq_par}
\end{equation}
provided each $f_\ell$ is continuous, regardless of whether the stages are run concurrently or sequentially.
\end{contractbox}

Viewed across all $L$ stages at once, the pipeline is a single Markov chain over the joint state $\mathbf{S} = (S^{(1)}, \dots, S^{(L)})$ formed by concatenating the states of the $L$ stages. Each stage advances its own component every cycle on the still-converging estimates of the stages feeding it, in the manner of a Jacobi-style fixed-point iteration over the whole unrolled chain. This is similar to the parallel-in-time relaxation of the Parareal family of evolution-equation solvers~\cite{lions2001parareal}, which iterate the entire time domain to a fixed point rather than marching step by step, and the same Picard-style parallelisation has recently been shown to draw samples from diffusion models in far fewer sequential rounds than their nominal step count~\cite{shih2023paradigms}. What relaxes is this joint state, and the end-to-end wall-clock is set by the time the joint chain takes to settle rather than by the sum of the per-stage latencies.

The contrast with conventional execution is the essential point. In a standard pipeline each stage must produce a finished, exact output before the next can consume it, so an $L$-stage computation costs the sum of the per-stage latencies. Here the stages settle together rather than in turn, so that at the accuracy the output requires the joint relaxation time can fall well below $L$ times a single stage. This is not to say that depth comes for free. Information must still propagate through the stages, so the advantage narrows as the target accuracy tightens.


\section{CN101: A Digital Thermodynamic Computer}
\label{sec:cn101}


Having established a formalism for equilibrium thermodynamic computing that is substrate-independent, we now get to the second core contribution of this work: realising it in digital silicon through the test chip CN101. The functions the chip is built to evaluate are deep nested compositions of affine maps and elementwise nonlinearities,
\begin{align}
  x^{(L)} = \sigma\!\Big(W^{(L)}\,\sigma\big(\cdots\,\sigma(W^{(1)} x + b^{(1)})\,\cdots\big) + b^{(L)}\Big),
\end{align}
a class we return to in \cref{sec:experiments}. This nesting and how the output of each layer is the input of the next, such that execution runs as a deep sequential path, is the core property we consider for the chip's functional design.

\subsection{Architecture}
CN101 is the first silicon instantiation of the \emph{Carnot Architecture}, a general design for realising the equilibration formalism of thermodynamic computing in digital logic through stochastic computing. In the following section we will discuss the basics of stochastic computing, the tile architecture representing individual stages of computation, how tiles are connected using a Stochastic Streaming Network-On-Chip (SSNoC) to compose stages, tighten the connection of the architecture and the equilibration formalism, and finally discuss how multiple chips can be combined to handle more compute intensive workloads.

\subsubsection{Stochastic computing}
\begin{figure}[t]
\centering
\begin{tikzpicture}[font=\footnotesize,
  cell/.style={draw=black!55,minimum size=4mm,inner sep=0,line width=0.3pt},
  on/.style={cell,fill=black!70,text=white},
  off/.style={cell,fill=black!5},
  lbl/.style={font=\scriptsize\itshape,text=black!55}]

\node[font=\bfseries,anchor=east] at (-0.9,3.5){(a)};
\node[lbl,anchor=west] at (0,3.98){Fixed-point: each bit a fixed place value};
\foreach \b/\i in {1/0,0/1,1/2,1/3}{
  \ifnum\b=1 \node[on] at (\i*0.5,3.5){1}; \else \node[off] at (\i*0.5,3.5){0};\fi}
\node[font=\scriptsize,text=black!50] at (0,3.12){$\tfrac12$};
\node[font=\scriptsize,text=black!50] at (0.5,3.12){$\tfrac14$};
\node[font=\scriptsize,text=black!50] at (1.0,3.12){$\tfrac18$};
\node[font=\scriptsize,text=black!50] at (1.5,3.12){$\tfrac{1}{16}$};
\node[anchor=west] at (2.05,3.5){$=\;0.6875$};
\node[lbl,anchor=west] at (0,2.68){Stochastic: value is the fraction of ones};
\foreach \b/\i in {0/0,1/1,1/2,0/3,1/4,0/5,1/6,1/7,0/8,0/9,1/10,0/11}{
  \ifnum\b=1 \node[on] at (\i*0.42,2.2){}; \else \node[off] at (\i*0.42,2.2){};\fi}
\node[anchor=west] at (5.05,2.2){$\hat{x}_T=\tfrac{6}{12}=0.5$};

\node[font=\bfseries,anchor=east] at (-0.9,1.0){(b)};
\begin{scope}[font=\scriptsize]
\draw[line width=0.5pt,fill=black!5] (0,0.65)--(0,1.35)--(0.35,1.35) arc(90:-90:0.35)--(0,0.65)--cycle;
\node at (0.3,1.0){\textsc{and}};
\draw[line width=0.4pt] (-0.55,1.15)--(0,1.15); \draw[line width=0.4pt] (-0.55,0.85)--(0,0.85);
\node[anchor=east] at (-0.55,1.15){$p$}; \node[anchor=east] at (-0.55,0.85){$q$};
\draw[line width=0.4pt] (0.7,1.0)--(1.1,1.0);
\node[anchor=west] at (1.1,1.0){mean $pq$ \;\textcolor{black!60}{\itshape multiply, $0.5\wedge0.4\to0.20$}};
\draw[line width=0.5pt,fill=black!5] (-0.1,-0.35) .. controls (0.13,0) .. (-0.1,0.35) .. controls (0.4,0.35) and (0.65,0.15) .. (0.83,0) .. controls (0.65,-0.15) and (0.4,-0.35) .. (-0.1,-0.35) -- cycle;
\node at (0.3,0){\textsc{or}};
\draw[line width=0.4pt] (-0.65,0.15)--(-0.02,0.15); \draw[line width=0.4pt] (-0.65,-0.15)--(-0.02,-0.15);
\node[anchor=east] at (-0.65,0.15){$p_1$}; \node[anchor=east] at (-0.65,-0.15){$p_2$};
\draw[line width=0.4pt] (0.83,0)--(1.23,0);
\node[anchor=west] at (1.23,0){mean $1-\prod_i(1-p_i)$ \;\textcolor{black!60}{\itshape accumulate}};
\end{scope}
\end{tikzpicture}
\caption{\textbf{Fixed-point versus stochastic representation.} \textbf{(a)}~A fixed-point word assigns each bit a fixed place value, whereas a stochastic bitstream encodes a value as its fraction of ones, recovered as the time-average $\hat{x}_T$. \textbf{(b)}~In this representation arithmetic reduces to single logic gates: an \textsc{and} gate multiplies two independent streams to mean $pq$ (worked example $0.5\wedge0.4\to0.20$), and an \textsc{or} gate accumulates a column of products. Note that this saturates to mean $1-\prod_i(1-p_i)$, which has to be corrected at readout.}
\label{fig:sc_repr}
\vspace{-.5em}
\end{figure}
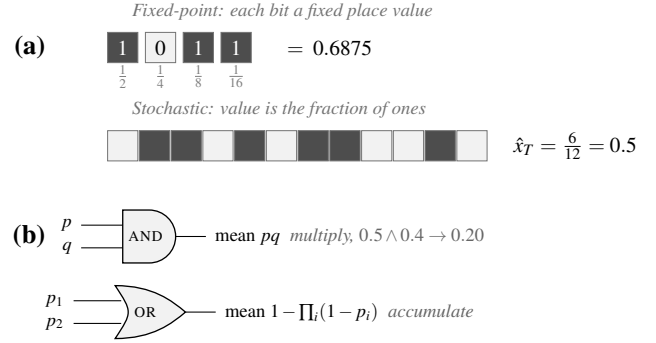

Rather than the binary word of a fixed-point representation, in which each bit carries a fixed place value, stochastic computing encodes a value in a random bitstream. A value $x$ is carried by a sequence of independent Bernoulli bits $b_1, b_2, \dots$ with $\mathbb{E}[b_t] = x$, and is recovered as the time-average
\begin{align}
  \hat{x}_T = \frac{1}{T}\sum_{t=1}^{T} b_t,
  \qquad
  \hat{x}_T \xrightarrow{\text{a.s.}} x,
  \qquad
  \mathrm{Var}(\hat{x}_T) = \frac{x(1-x)}{T}.
\end{align}
The encoded value is therefore an expectation and the readout an unbiased estimate of it. Reading a value as the time-average of an accumulator driven by a stochastic bitstream is the same stochastic-integration mechanism prior stochastic-computing hardware uses to solve differential equations~\cite{liuHan2017,liuGrossHan2020}. Precision, similar to the thermodynamic computing framework outlined before, is set by the length of the averaging window rather than by a fixed word width (\cref{fig:sc_repr}). Every bit in the stream carries equal weight, unlike the graded place values of a fixed-point or floating-point word, so that reaching a precision $\epsilon$ costs of order $\epsilon^{-2}$ bits. As such,  the representation is cheap when a coarse estimate suffices and expensive when an exact one is demanded. Similar to the anytime precision property discussed earlier.

Specifically, CN101 supports unipolar as well as \emph{split-unipolar} encoding, the latter of which is used for the experiments reported in the remainder of this paper. In split-unipolar encoding a single unipolar stream represents a value in $[0,1]$ and a signed value is carried by a pair of such streams, so that signed arithmetic reduces to unipolar arithmetic on two channels~\cite{acoustic}.

\begin{figure*}[t!]
  \begin{minipage}[t]{0.70\linewidth}
    \centering
    \includegraphics[width=\linewidth]{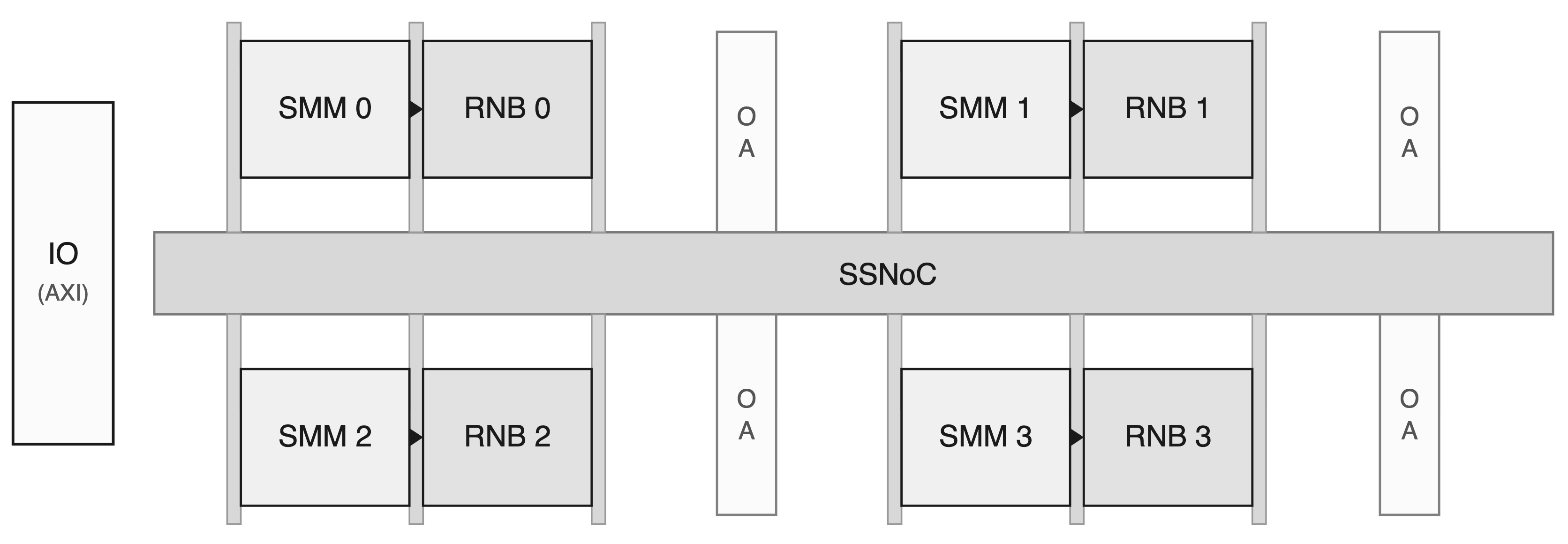}
    \subcaption{}
  \end{minipage}%
  \hfill
  \begin{minipage}[t]{0.3\linewidth}
    \centering
    \includegraphics[width=0.78\linewidth]{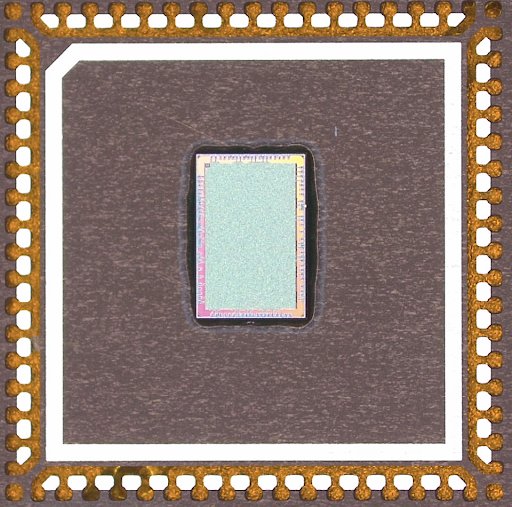}
    \subcaption{}
  \end{minipage}

  \begin{minipage}[t]{0.50\linewidth}
    \centering
    \includegraphics[width=\linewidth]{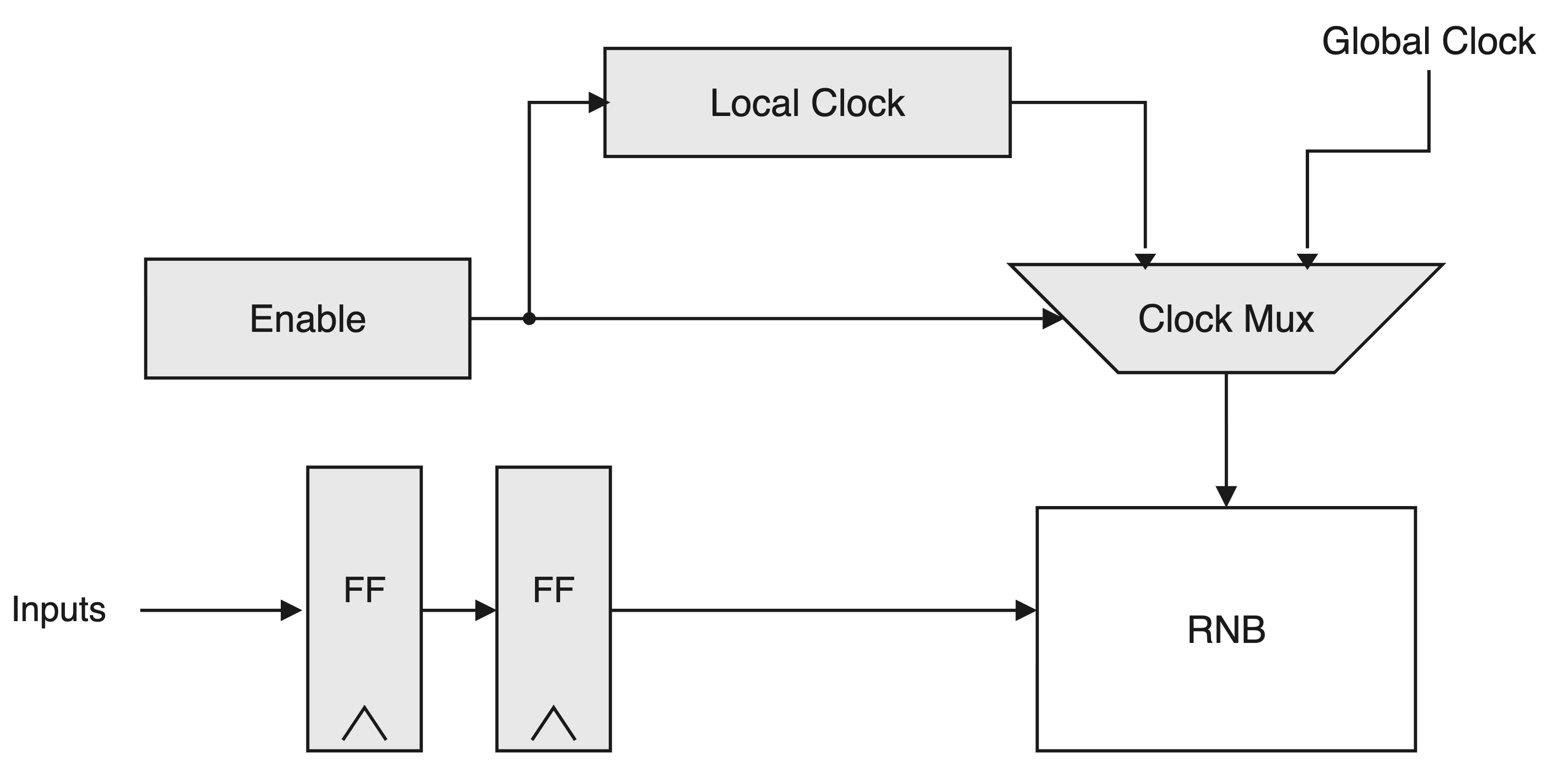}
    \subcaption{}
  \end{minipage}%
  \hfill
  \begin{minipage}[t]{0.46\linewidth}
    \centering
    \includegraphics[width=0.75\linewidth]{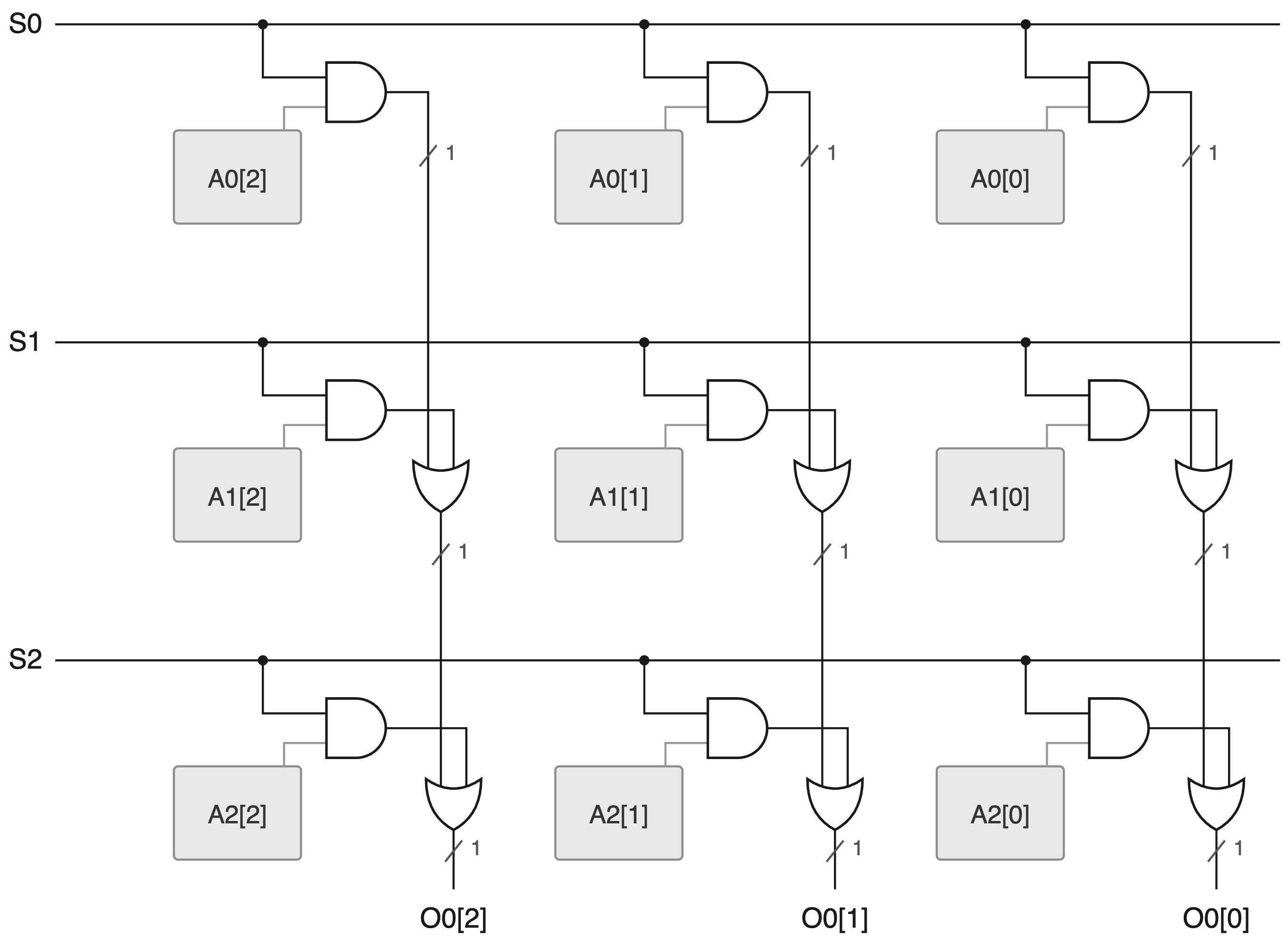}
    \subcaption{}
  \end{minipage}

  \caption{\textbf{CN101 stochastic digital thermodynamic computing.}
    \textbf{(a)}~Top-level architecture. Four tiles (each comprising an SMM, RNB, and OA) are connected by the SSNoC. The IO block provides host configuration access.
    \textbf{(b)}~Die photograph of CN101.
    \textbf{(c)}~Polysynchronous clocking. Each RNB is driven by either a global clock tree or a local ring oscillator, selected by a clock mux; two flip-flops handle metastability at the clock domain boundary.
    \textbf{(d)}~OR-based accumulation in the stochastic matrix multiplier with AND gates computing bitstream products per element.
    }
  \label{fig:chip}
  \vspace{-1.1em}
\end{figure*}

For two independent unipolar streams with means $p$ and $q$, the bitwise \textsc{and} has mean $pq$, so that multiplication is a single \textsc{and} gate rather than a full multiplier. Similarly, summation is implemented using an \textsc{or} gate~\cite{gaines1969,alaghi2013}. As such, a matrix--vector product can be implemented efficiently using a grid of \textsc{and} gates feeding column-wise \textsc{or} accumulation. This forms the basis of the stochastic matrix multiplier (SMM) (\cref{fig:chip}d) at the core of each tile. This \textsc{and}-multiply, \textsc{or}-accumulate datapath is often used in stochastic-computing neural-network accelerators~\cite{ren2017scdcnn,kim2016dynamic,sim2017scacc}.

Specifically, for the SMM on CN101, the weight operand $W$ is held as a deterministic 8-bit value rather than as a second stream, so that only the input $x$ adds variance. Each input bit is gated against the eight weight bits separately, so that every bit plane of $W$ carries its own \textsc{or} accumulation column, and the eight columns are recombined by place value at readout. The \textsc{and}/\textsc{or} datapath is in this way kept intact while the weight retains its precision, which is why the array delivers a multi-bit result.

\subsubsection{The tile}
One SMM together with a reconfigurable neuron bank (RNB) form the core computation unit on CN101, tiles. Each tile evaluates a single layer $\sigma(Wx + b)$ in full. The term \emph{tile} is used in this specific sense throughout: a tile is a complete affine-plus-nonlinear step, not a sub-block of one matrix multiplication partitioned for data locality, as \emph{tile} denotes on a GPU. The SMM stores the weight matrix $W$ and forms $Wx$ through the \textsc{and}/\textsc{or} arithmetic above. The RNB adds the bias $b$ and applies the nonlinearity $\sigma$. Each CN101 contains 4 of these tiles (\cref{fig:chip}a).

The nonlinearity is implemented along one of two paths. Where possible, we rely primarily on ReLU activations, which are trivial in the split-unipolar encoding since they act separately on the two sign channels. However, for generic nonlinearities the RNB also has access to a finite-state machine, the standard construction for nonlinear functions in stochastic computing~\cite{brownCard2001,li2020fsm}.

Within a tile the SMM passes its multi-bit result to the RNB over a direct, deterministic path, so that applying a layer adds no sampling noise of its own. The inputs $x$ are themselves stochastic estimates, so that the tile's output is still a stochastic quantity, with variance inherited from its inputs. What the deterministic intra-tile path avoids is the additional variance that re-encoding the value into a fresh bitstream would introduce.

Beyond the multiplier and the neuron bank, each tile carries an output accumulator (OA), a readout channel whose role in composition we take up in \cref{sec:composition}.

\subsubsection{Composing tiles}
With a single tile evaluating a single layer, a deeper model can be assembled by composing multiple tiles. This is achieved by passing the output of one tile to the input of the next. CN101 carries this composition over its stochastic streaming network-on-chip (SSNoC), a reconfigurable Network-On-Chip that routes bitstreams between tiles. The bitstream representation required is generated within the RNB itself. Here each neuron carries a 65-bit linear-feedback shift register (LFSR) with a nonlinear output stage, of period exceeding $10^{19}$, so that this per-neuron pseudo-random number generator (PRNG) supplies the Bernoulli bits for that neuron's streams~\cite{neugebauer}.

As such, there exist two communication paths with complementary roles in the chip. Inside a tile, the SMM-to-RNB path is deterministic and multi-bit, so that a single layer is applied without added sampling noise. Conversely, between tiles, the SSNoC path carries bitstreams, is reconfigurable, and crosses between independent clock domains. The first preserves the accuracy of a layer and the second is what makes the composition of the chip into distinct equilibration units.

One specific property that stochastic communication between tiles allows is a clean separation between clock domains that requires no synchronisation, resulting in a globally-asynchronous locally-synchronous design~\citep{chapiro1984} at the level of a tile. A bitstream carries an expected value, and every bit of it carries the same expected value, so that the quantity a downstream tile estimates does not depend on which bits arrive or on exactly when they arrive. A delayed or dropped bit changes the number of samples averaged, not the value being averaged. As such, the SSNoC can move streams between tiles running on independent clocks without a global clock, each tile driven either by a global clock tree or by a local ring oscillator and left to settle at its own rate.

\cref{tab:chip} lists the physical parameters of CN101 (\cref{fig:chip}b).

\begin{table}[t]
  \centering
  \small
  \caption{Key physical parameters of CN101.
  }
  \vspace{-0.2cm}
  \label{tab:chip}
  \begin{tabularx}{\linewidth}{@{}lX@{}}
    \toprule
    \textbf{Parameter} & \textbf{Value} \\
    \midrule
    Tile count                & 4 \\
    SMM array (per tile)      & $64 \times 64$ \\
    Neurons per RNB           & 64 \\
    Total MAC cells           & 16{,}384 \\
    Total neuron accumulators & 256 \\
    Accumulator width         & 32 bits \\
    PRNG                      & 65-bit LFSR + nonlinear output stage \\
    Operating frequency       & $\approx$ 500 MHz \\
    \bottomrule
  \end{tabularx}
  \vspace{-0.5cm}
\end{table}

\subsection{CN101 and the equilibration formalism}
\label{sec:formalism_mapping}

Having described how CN101 evaluates a function, we now make its connection to the equilibration formalism precise. Crucially, the chip does not compute its outputs in closed form: each value it carries is the expectation of a bitstream, obtained as a time-average over a window of $T$ cycles, and the bitstreams passing through the active tiles together form a single ergodic process. The quantity the chip reports is the long-run behaviour of that process rather than the state it occupies on any one cycle, so that the chip realises a function in the precise sense of \cref{sec:framework}, as the stationary expectation of its own dynamics.

This joint relaxation is distinct from ancestral sampling. The tiles do not draw a finished sample $x^{(\ell)}$ from a conditional $p(x^{(\ell)}\mid x^{(\ell-1)})$ and hand it downstream. Instead, each updates every cycle on the current, still-converging estimate of the tile feeding it, realising the joint fixed-point relaxation of \cref{sec:scaling}~\cite{lions2001parareal,shih2023paradigms}. What the chip reads out is the mean of that process, the layer output $\sigma(Wx+b)$ recovered as a time-average, and never a single draw from it.

The four ingredients of the formalism each have a counterpart on the chip. The state $S_t$ is the joint configuration of the accumulators across the active tiles, an integer vector $S_t \in \mathbb{Z}^d$ whose dimension $d$ is the number of active neurons in the RNBs. The generator $\Lstar$ advances this state by one cycle under the pseudo-random bits the neuron banks supply, and because the state space is discrete it is a Markov chain rather than the continuous Langevin operator that \cref{sec:framework} admits as its analogue counterpart. The weights, biases, and activations loaded onto the chip fix the stationary distribution $\pi$ satisfying $\Lstar\pi = 0$, while the readout $\varphi$ is taken from the output stream, whose time-average
\begin{align}
  \hat{y}_T = \frac{1}{T}\sum_{t=1}^{T} \varphi(S_t)
\end{align}
converges to $\Epi[\varphi(S)]$ as the window grows. As such, the function the chip computes is $f(x) = \Epi[\varphi(S)]$, the definition given in \cref{sec:framework}.

The same correspondence shows how CN101 meets the three challenges of \cref{sec:challenges}. Implementation is the most direct: the accumulator dynamics are the generator, carried out by the chip's own logic rather than simulated on a host. Encoding is solved by the construction of tiles: instead of searching for a single generator whose stationary expectation is an arbitrary $f$, CN101 builds $f$ layer by layer, each tile realising one affine-plus-nonlinear step as the stationary expectation of its own stream, with the weights supplied by ordinary training. What remains is efficiency, the requirement that the dynamics mix fast enough for $\hat{y}_T$ to converge in useful time; this is a quantitative property of the chip rather than a question of principle, and we characterise it directly in \cref{sec:validation}.

\subsection{Composition across tiles and chips}
\label{sec:composition}

To realise more complex functions, we now turn to how the design of CN101 supports composition over multiple chips. Both directions of scaling, depth and width, exploit the same feature of how a tile computes: a tile does not deliver a value at a fixed moment, but produces an estimate that sharpens the longer it runs. A composition of tiles then depends only on these estimates, and not on the tiles sharing a synchronised state, with two consequences. First, tiles may run on independent clocks, since combining them requires no shared notion of when a value is ready. Second, and more important for scaling, a downstream tile may consume an upstream estimate that has not yet converged, so that all stages advance together rather than in turn.

Taken together, these properties make the chip boundary a matter of implementation rather than of principle. Within a chip the tiles couple directly over the SSNoC, exchanging streams as they relax on independent clocks. Across a chip boundary the same coupling is realised by exchanging their running estimates, so that many chips act as a single system of tiles converging in unison to the collective result (\cref{app:multichip}). We examine this scaling along the two axes in turn, each applying across chips as readily as within one.

\paragraph{Depth: composing chips in sequence.}
The first axis adds depth by spreading a deep model across several chips. Writing the model as the composition $f = f_L \circ \cdots \circ f_1$ of its layers, we partition the layers into contiguous blocks, one block to a chip, so that the output of one chip is the input of the next. Each chip holds a running estimate of its block's output, which the chip downstream consumes. That the chain converges to the intended result is the sequential-parallelism property of \cref{sec:scaling}: provided each layer map $f_\ell$ is continuous in its input,
\begin{align}
  \hat{y}_T^{(L)} \;\xrightarrow[T\to\infty]{}\; \big(f_L \circ \cdots \circ f_1\big)(x).
\end{align}

\paragraph{Width: composing tiles into a larger matrix.}
The second axis adds width by combining tiles, on one chip or across several, whose weight matrices together represent a matrix larger than a single $64 \times 64$ array can hold. This is similar to the tiling used for dense linear algebra on a GPU. Partitioning the weight matrix into blocks $W_{ij}$ and the input correspondingly into blocks $x_j$, each tile evaluates one block product $W_{ij}\,x_j$, and the $i$-th block of the layer output is the sum
\begin{align}
  (Wx)_i = \sum_j W_{ij}\, x_j
\end{align}
of the partial products of the tiles assigned to it. Crucially, the tiles equilibrate independently, so that at any instant their partial products have converged to different degrees. This, however, does not bias the result, since each is an estimate of its block product and, expectation being linear, their sum is an estimate of $(Wx)_i$ whatever the individual states. The unevenness in convergence across tiles enters only the variance of the combined estimate and not its expectation.

\subsubsection{Output Accumulator}
Both axes rest on the same concept: a converging estimate is preserved under the summation that builds width and under the continuous composition that builds depth. What this earns in practice is freedom in how, and how often, the estimates are read from the tiles and passed on.

To support this communication across chips each tile carries an \textit{Output Accumulator} (OA). Crucially, the OA performs no part of the tile's computation; it is a dedicated, non-disruptive channel that taps the tile's output as the stream passes onto the SSNoC and integrates it into a running sum. The OA allows for the current time-averaged estimate to be read at any moment, without halting the neuron bank or perturbing the dynamics that generate the stream. A downstream stage, located on a different chip, can then draw its input from this accumulated value rather than from the live neuron bank.

The reads from these accumulators can be staggered across tiles and taken at different rates. Because a downstream tile tolerates a stale input, a bounded delay between when an estimate is produced and when it is consumed changes only the transient accuracy of the downstream stage, and not the value it converges to. In place of a global barrier, each downstream input is refreshed over a window of cycles from the current reading of its source, so that no two inputs are updated at the same instant and no stage waits on another. Within a chip this exchange is continuous, the streams flowing between tiles as they run. Across chips it proceeds window by window, the window length setting how closely the assembly approaches the fully concurrent limit.


\section{Single-chip validation: Digital TLA and Variational Autoencoders}
\label{sec:validation}
Having described how CN101 evaluates and composes functions, we now validate that it does so correctly using a single chip. We use two different workloads for this. The first is a linear system, the problem class for which the equilibration formalism was originally written down and on which the chip's output can be checked against an exact solution. The second is a small generative model, a conditional variational autoencoder, whose output is checked against a floating-point reference of the same trained network while it executes the full multi-layer cascade.


\begin{figure*}[t!]
  \includegraphics[width=\linewidth]{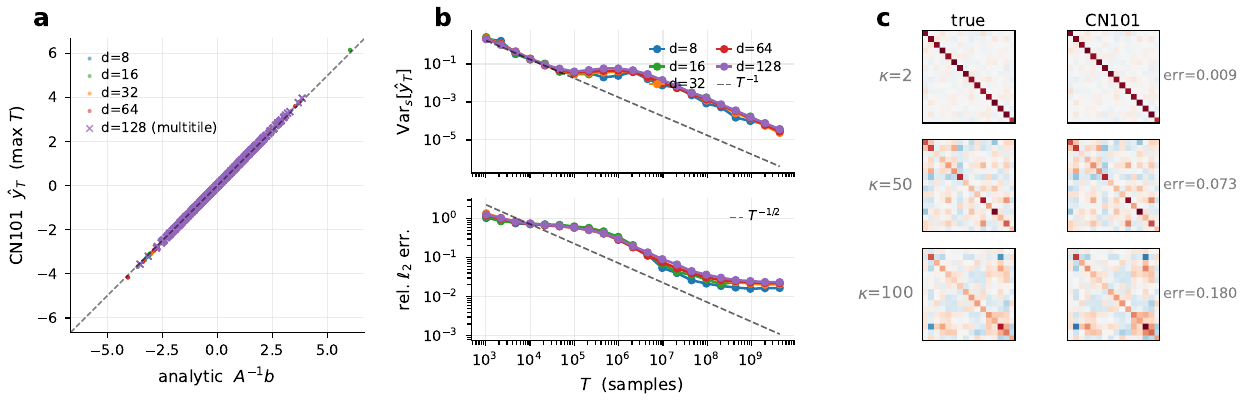}
  \caption{\textbf{Thermodynamic linear algebra on CN101.}
    \textbf{(a)}~Parity plot of CN101 $\hat{y}_T$ vs analytic $A^{-1}b$ across dimensions $d \in \{8, 16, 32, 64\}$ on a single tile and $d = 128$ using all four tiles in multi-tile composition (shown as $\times$). All points lie on the diagonal.
    \textbf{(b)}~Variance (top) and relative $\ell_2$ error (bottom) as functions of $T$, with $T^{-1}$ and $T^{-1/2}$ reference lines. Each curve averages 16 trajectories (4 system matrices $\times$ 4 PRNG seeds per matrix) per dimension.
    \textbf{(c)}~Ground-truth (left) and CN101-reconstructed (right) $A^{-1}$ at $d = 16$ for $\kappa \in \{2, 50, 100\}$.}
  \label{fig:tla}
\end{figure*}

\subsection{Digital thermodynamic linear algebra}
\label{sec:tla}

We begin with linear systems $Ax = b$ with symmetric positive-definite $A$, realised on the chip as thermodynamic linear algebra~\cite{aifer2024_TLA,melanson2025}, for which the encoding is analytically controlled. It is a deliberately narrow validation rather than a representative run: it checks correctness against a known answer on a different datapath from the stochastic-computing equilibration the generative models use. The correct stationary distribution is known exactly, so any deviation between the chip's output and $A^{-1}b$ would directly be attributable to implementation error.

For this workload every tile's activation is set to linear integration, which implements the lattice random-walk discretisation of the Ornstein--Uhlenbeck dynamics~\cite{duffield2025lattice, mensch2026robust}: integer-valued accumulator updates whose stationary expectation is $A^{-1}b$ to $O(\Delta t)$. The weight matrices encode $A$ and $b$ rather than a trained network. Everything else (PRNG seeds, routing tables, and the window length $T$) is configured as for any other workload.

\subsubsection{Results}
We characterise three properties in turn: correctness of the solution, $T^{-1}$ variance scaling, and the operating envelope set by problem conditioning.

\paragraph{Correctness.}
\cref{fig:tla}a shows a parity plot of every component of the chip's time-averaged readout $\hat{y}_T$ at maximum $T$ against the corresponding component of the analytic solution $A^{-1}b$, across dimensions $d \in \{8, 16, 32, 64\}$ on a single tile and $d = 128$ using all four tiles composed via the SSNoC. All points lie on the diagonal across the full dynamic range. The multi-tile result is indistinguishable from the single-tile results, confirming that the SSNoC composition preserves the formulation without introducing systematic error.

\paragraph{Convergence rate.}
\cref{fig:tla}b shows variance and relative $\ell_2$ error as functions of $T$ over six orders of magnitude. The top panel plots $\mathrm{Var}_s(\hat{y}_T)$ across 16 trajectories per dimension (4 system matrices $\times$ 4 independent PRNG seeds per matrix); the bottom panel plots the relative error for a representative seed. Both panels show a consistent three-regime structure across all dimensions. In the early regime ($T \lesssim 10^5$), variance decays at slope $\approx -1$, matching the $T^{-1}$ prediction, while the error decreases slowly as the chain is still shedding memory of its initial state; for $T \lesssim \trelax$ this is the burn-in of \cref{sec:challenges}, in which the initial-state bias rather than the variance dominates the error. In the mid regime ($10^5 \lesssim T \lesssim 10^7$) the variance decay visibly slows as the dynamics enter a transitional mixing phase; correspondingly, the error curve steepens towards its $T^{-1/2}$ descent as variance becomes the dominant term in the MSE. In the late regime ($T \gtrsim 10^7$), variance resumes its $T^{-1}$ slope before all curves flatten onto the bias floor at relative error $\approx 10^{-2}$, consistent with finite weight-register precision in the SMM. The convergence curves across dimensions are nearly coincident, indicating that the per-tile correlation time $\tcorr$, which sets the rate of the variance-limited mixing regime, is not strongly sensitive to problem dimension in this regime.

\paragraph{Conditioning.}
\cref{fig:tla}c shows the reconstructed inverse $A^{-1}$ at $d = 16$ for three condition numbers, assembled column-by-column. At $\kappa = 2$ the reconstruction is visually indistinguishable from the ground truth (relative error 0.009). At $\kappa = 50$ the structure is accurately recovered (relative error 0.073). At $\kappa = 100$ the main structure remains visible with some residual error (relative error 0.180), establishing the operating envelope of the current generation at this problem size.

\begin{figure*}[b!]
  \includegraphics[width=\linewidth]{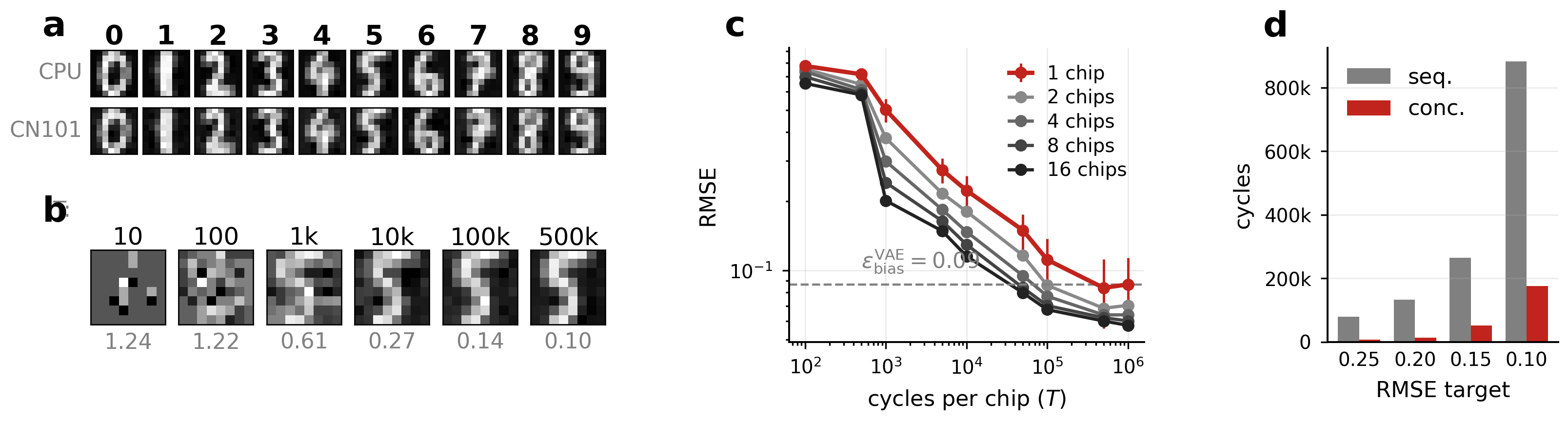}
  \caption{\textbf{Conditional MNIST generation on a single chip.}
    \textbf{(a)}~Reconstructions of all ten MNIST classes at the concurrent operating point ($T{=}500\text{k}$ cycles). Top row, CPU floating-point reference, bottom row, CN101.
    \textbf{(b)}~Progression of a single digit's reconstruction as the cycle budget $T$ increases from $10$ to $500\text{k}$; values below each panel are the scale-invariant RMSE against the CPU reference.
    \textbf{(c)}~RMSE vs.\ $T$. Solid line: single-chip result at the per-$T$ optimal operating point. Shaded lines: mean of $K$ independent readouts (distinct RNG seeds), $K \in \{2, 4, 8, 16\}$. Dashed line: bias floor $\epsilon_\mathrm{bias}^\mathrm{VAE}$.
    \textbf{(d)}~Cycles required to reach each target RMSE $\tau$ for the sequential ($\sum_i T_i$) and concurrent ($T$) configurations.}
  \label{fig:vae}
\end{figure*}

\subsection{Conditional MNIST generation with a single-chip VAE}\label{sec:vae}

The second validation workload is a conditional variational autoencoder (VAE)~\cite{kingma2014vae,sohn2015cvae}.
A conditional VAE on $8{\times}8$ MNIST fits on a single CN101, its four decoder layers mapping one-to-one onto the four tiles, and as such provides the within-chip demonstration of parallel sample aggregation and sequential parallelism. Unlike the linear system, the intended output here is not known in closed form; we validate instead against a floating-point reference of the same trained model.

\subsubsection{Model architecture and training}

The decoder uses four fully-connected layers of width 64. The first layer concatenates a 16-dimensional latent code with a 10-way one-hot class label and projects to a 64-dimensional hidden representation. Two further $64{\times}64$ hidden layers with ReLU activations follow, and the output layer projects to a 64-dimensional image vector reshaped to $8{\times}8$. The network is trained on $8{\times}8$ MNIST-like handwritten digits captured natively at $8{\times}8$~\cite{alpaydin1998digits} rather than downsampled from the $28{\times}28$ MNIST~\cite{lecun1998mnist} they resemble, using the reparametrisation trick and a class-conditional evidence lower bound (ELBO) objective; training runs in floating point on a CPU and the weights are then loaded onto CN101.

The four layers map one-to-one onto the four tiles: one weight matrix per tile, with the per-tile activation register set to ReLU for the three hidden layers and to LINEAR for the output layer. Two execution configurations are compared: concurrent and sequential. In the \emph{concurrent} configuration all four tiles execute simultaneously and bitstreams flow between them through the network-on-chip. In the \emph{sequential} configuration each layer runs in isolation to convergence and the integer output of each stage is re-encoded and fed into the next. The final tile's output is the stream consumed by the output accumulator.

\subsubsection{Results}

\cref{fig:vae}a shows CN101 reconstructions of all ten digit classes alongside the floating-point reference at the concurrent operating point ($T = 500{,}000$ cycles per tile; hyperparameters in \cref{app:vae}). The reconstructed digits are recognisable across all ten classes with the scale-invariant RMSE averaged across 10 digits $\times$ 10 latent samples being $0.083 \pm 0.028$. \cref{fig:vae}b shows the evolution of a single digit's reconstruction as $T$ increases: at $T = 10$ the output is dominated by quantisation noise; the digit is clearly recognisable by $T = 10{,}000$ and converges onto the $T = 500{,}000$ limit by $T \approx 10^5$.

\paragraph{Anytime-precision.} \cref{fig:vae}c plots the RMSE as a function of $T$, with per-$T$ operating points selected by a search over the chip's scale and precision knobs (details in \cref{app:vae}). The curve drops from $\mathrm{RMSE} \approx 0.79$ at $T=100$ to a bias floor $\epsilon_\mathrm{bias}^\mathrm{VAE} \approx 0.09$ at $T = 10^6$, set by weight quantisation in the multiplier register banks and independent of $T$. At this operating point the reconstructed digits are perceptually indistinguishable from the floating-point reference at the output resolution of $8\times8$ pixels. Reducing the floor further would require higher-precision weight registers. This sweep over $T$ is an example of the anytime-precision axis discussed in \cref{sec:scaling} on a generative workload.

\paragraph{Parallel sample aggregation.}
\cref{fig:vae}c overlays four curves obtained by averaging across $K \in \{2, 4, 8, 16\}$ independent RNG seeds at each $T$. The $K$ curves separate, approaching the $1/\sqrt{K}$ spacing of the variance-limited regime (at $T = 1000$ the $1{\to}16$ seed ratio is $2.3\times$, short of the ideal $4\times$ because the bias floor still contributes). At small $T$ the burn-in transient dominates the error and, being shared by replicas started alike, is not reduced by averaging (at $T = 100$ the ratio is $1.2\times$). At large $T$ the bias floor limits further convergence.

\paragraph{Sequential parallelism.}
Sequential parallelism~\cref{eq:seq_par} predicts that the concurrent configuration delivers the same output as the sequential configuration, but with a joint relaxation time rather than the sum $\sum_i T_i$ of the per-stage times. \cref{fig:vae}d quantifies this at four RMSE targets $\tau \in \{0.25, 0.20, 0.15, 0.10\}$. For each target the smallest cycle budget is found for both execution forms: concurrent and sequential (search details in \cref{app:vae}). The concurrent budget grows from $7{,}300$ cycles at $\mathrm{RMSE} = 0.25$ to $175{,}300$ cycles at $\mathrm{RMSE} = 0.10$. The sequential total $\sum_i T_i$ tracks the same slope but is larger by a factor of $5$--$11\times$; at $\mathrm{RMSE} = 0.10$ the concurrent configuration reaches the target in $1.75 \times 10^5$ cycles against $8.8 \times 10^5$ cycles for sequential.


\vspace{0.3cm}
\section{Thermodynamic computing for generative AI}
\vspace{0.2cm}
\label{sec:experiments}
Having validated the single-chip operation of CN101, we now focus on its core target workload: generative AI. Modern Generative AI approaches are inherently sequential in nature and therefore benefit most from the scaling axis of thermodynamic computing implemented on CN101. Progress in deep learning has repeatedly come from depth rather than width: the breakthrough on ImageNet came not from wider networks but from very deep residual networks that made optimisation at depth stable~\cite{he2016resnet}. The generative models that followed inherited this character by construction. Diffusion and flow-matching models are defined as continuous-time processes and generated by integrating a long chain of dependent steps, and large language models generate one token at a time, each conditioned on every token before it. Sequential depth is therefore not incidental to these models; it is where their capability comes from.

The standard hardware these models run on, however, rewards the opposite. The parallelism of the GPU favours width over depth, larger batches and more parallel lanes over long dependent chains, and this ``hardware lottery''~\cite{hooker2021lottery} has shaped a decade of model design towards what existing accelerators execute cheaply, at the expense of the sequential depth on which the most capable generative models depend. Models have been made wider and shallower, and batched ever more aggressively, to suit the hardware rather than the structure of the problem.

The scaling axes of \cref{sec:scaling}, and sequential parallelism in particular, invert this trade-off. Because a downstream stage can consume an upstream estimate while it is still converging, dependent stages relax concurrently rather than in sequence. Where the intermediate stages tolerate imprecision a deep inherently sequential computation can complete in close to the time of a single stage rather than in proportion to its depth, as we will experimentally validate in the coming sections. As such, thermodynamic computing matches the sequential structure of modern generative models rather than working against it.

To be able to demonstrate this, each workload in this section runs across six CN101 chips that realise the multi-chip composition of \cref{sec:composition}, each holding a block of the model and advancing on the still-converging estimates of the blocks upstream of it. The present chip is a prototype, and composing six of them is in part how we assemble the weight capacity these models need; but it is equally a benchmark of the composition itself, which is the mechanism by which larger models are assembled on future hardware.

\subsection{Generative models as deep nested maps}
\label{sec:ode_sde}

Before continuing, we now first make explicit that modern generative methods, such as diffusion~\cite{ho2020ddpm,sohldickstein2015} and flow matching, can be described as a deep nesting of affine maps and nonlinearities of the form $\sigma(Wx+b)$ and therefore map well to the core operating model of thermodynamic computing.

A diffusion or flow-matching model generates by integrating an ordinary differential equation $\dot{x} = v_\theta(x, t)$ (or equivalent stochastic differential equation \cite{duffield2026complete}). Its Euler discretisation
\begin{align}
  x_{k+1} = x_k + \Delta t\, v_\theta(x_k, t_k)
\end{align}
is a residual connection, so that the unrolled integration is a deep residual network in which the learned field $v_\theta$ is the per-step layer~\cite{he2016resnet,chen2018node,weinan2017}. The number of these steps is the Number of Function Evaluations (NFE), which we denote $N$. A network of $L$ layers unrolled over $N$ steps is a nested computation $L\times N$ layers deep, and it is this product, not $L$ alone, that sets the sequential length of generation.

The stochastic case extends this without leaving the nested form. A discretised stochastic differential equation, whether the noise enters additively, multiplicatively, or through the argument of the drift, takes the Euler--Maruyama form~\cite{kloeden1992}
\begin{align}
  x_{k+1} = x_k + f(x_k,\, \Delta W_k), \qquad \Delta W_k \sim \mathcal{N}(0, \Delta t).
\end{align}
Once the noise increments $\Delta W_k$ and the step schedule are fixed, the unrolled integration is again a deterministic deep nested map, differing from the ODE case only in the injected randomness~\cite{song2021sde}.

%

\subsection{CIFAR-10 generation with a convolutional flow}
\label{sec:cifar}

Flow matching~\cite{lipman2023} is one of the core methods of modern generative AI. It trains a velocity field so that integrating it transports a simple prior into the data distribution. Flow models built this way underlie much of image, video and molecular generation. It is therefore our first benchmark of what thermodynamic computing offers generative AI. We train a convolutional flow-matching model to generate CIFAR-10 images~\cite{krizhevsky2009cifar} and run it across the CN101 chips, asking two things: whether the substrate reproduces the same results in expectation as when run on a traditional accelerator, and how the thermodynamic computing scaling axes act on the sequential structure of the generation.

\subsubsection{Setup}
The prototype chip's only operation is a $64\times64$ matrix multiply with no direct native support for convolutions. As such, we express each convolution as a host-side \emph{shift}~\cite{wu2018shift} followed by an on-chip \emph{$1\times1$ convolution}. The shift re-indexes each channel's spatial map so that a pixel's neighbours are gathered into its own channel vector, a free memory operation, after which the $1\times1$ convolution, a per-pixel $\mathrm{Linear}(C,C)$, is exactly the chip's matmul, streamed over every pixel of the image (weight-stationary). The group normalisations and nonlinearities run on the host. The resulting model, \textsc{ChipUNetTiny}, is a ten-block U-Net~\cite{ronneberger2015} (a stem, two downsample--encoder stages, a bottleneck, two upsample--decoder stages with skip connections, and a head) on $32\times32\times3$ images. Its ten $1\times1$ blocks are pinned as ten weight-stationary tiles across the six-chip system (\cref{sec:composition}), with two further tiles reallocated to share the busiest layers' pixel streams, so every weight is resident throughout generation.

The learned velocity field is integrated over $N$ steps, each a full pass through the ten-layer U-Net, so the unrolled integration is $10\times N$ layers deep. The chip does not give each of those layers its own tile. The ten weight matrices are pinned one to a tile across the six CN101 chips, and because they recur at every integration step those ten tiles are reused, one evaluating each layer that shares its weights. This reuse bounds throughput, but it does not serialise the relaxation: sequential parallelism rests on the consistent-estimate property, so layers that share a tile still relax concurrently from one another's running estimates.

We train the flow-matching model on a set of ten CIFAR-10 images and apply reflow (rectified flow)~\cite{liu2022rectified} so that high-quality samples are produced in few integration steps. We use $N=10$ throughout, so the unrolled integration is 100 layers deep. The model is deliberately narrow in scope: it is overfit to these ten images and does not generalise, the prototype's on-chip weight memory being too small to hold a general CIFAR model. Every per-step matmul is read out as a time-average over a schedule rising to $5\times10^5$ cycles per tile. The first windows, over which the unrolled relaxation fills, are discarded from this average (a burn-in). We compare the chip-generated images against the floating-point reference of the same trained model.

\begin{figure*}[t!]
  \centering
  \includegraphics[width=\linewidth]{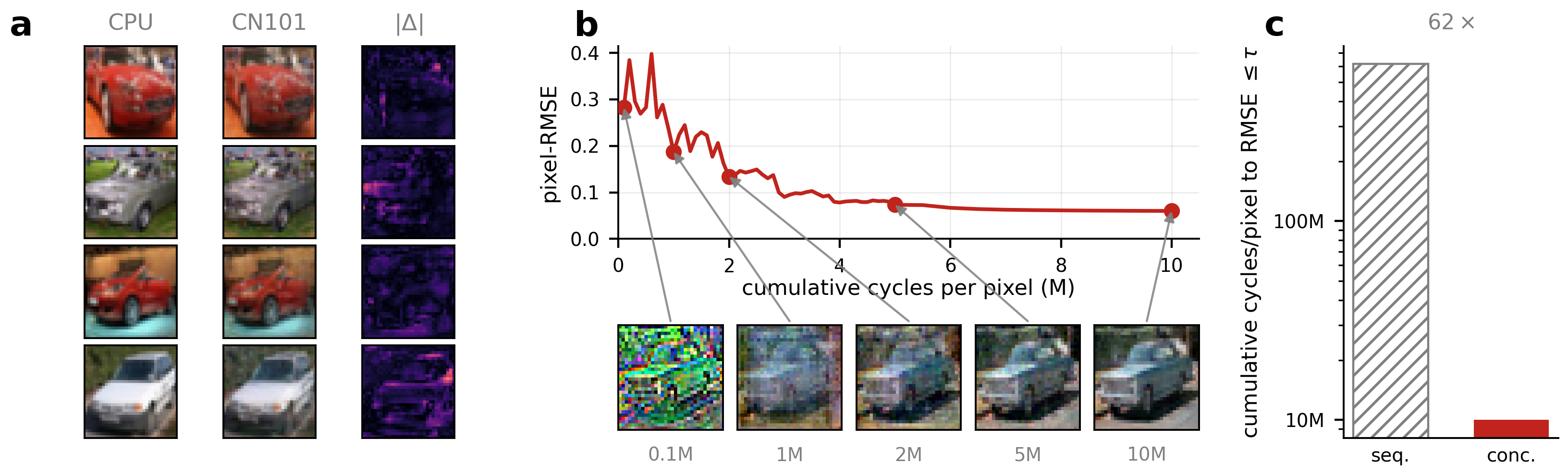}
  \caption{\textbf{CIFAR-10 generation on chip.} \textbf{(a)}~Four generated samples: the floating-point reference (CPU), the chip output (CN101), and their per-pixel absolute difference ($|\Delta|$).
  \textbf{(b)}~Pixel-RMSE of a single generated image against the reference as the cumulative cycle budget grows to 10M cycles per pixel, with the chip image at cumulative budgets of 0.1, 1, 2, 5 and 10M cycles below.
  \textbf{(c)}~Cumulative cycles per pixel to reach the same image quality ($\mathrm{RMSE}\approx 0.06$) for the 100-layer integration.
  }
  \label{fig:cifar}

  \vspace{-1em}
\end{figure*}

\subsubsection{Results}
The model generates recognisable CIFAR images, and the chip reproduces the floating-point reference of the same model, sample for sample, at a pixel-RMSE of $\approx 0.06$, with the residual concentrated in high-frequency detail rather than the composition (\cref{fig:cifar}a). A sample begins as quantisation noise and sharpens into a recognisable car as cycles accumulate, the pixel-RMSE falling over the schedule until it settles onto the floor set by the precision of the weight registers (\cref{fig:cifar}b). This is an example of the anytime-precision axis discussed in \cref{sec:scaling}, realised across multiple chips.

Comparing sequential and concurrent execution: run sequentially, each layer must converge before the next can use it, so the cost grows in proportion to the depth. Relaxed concurrently, the layers converge jointly, reaching the same image (pixel-RMSE $\approx 0.06$) in ${\approx}10$M cycles per pixel against the sequential run's ${\approx}620$M, some $62\times$ fewer at this target (\cref{fig:cifar}c). The sequential cost is a conservative bound: 60 of its 100 layers never meet the per-layer convergence target and exhaust the cycle cap.

\begin{figure}[t!]
  \centering
  \includegraphics[width=0.7\columnwidth]{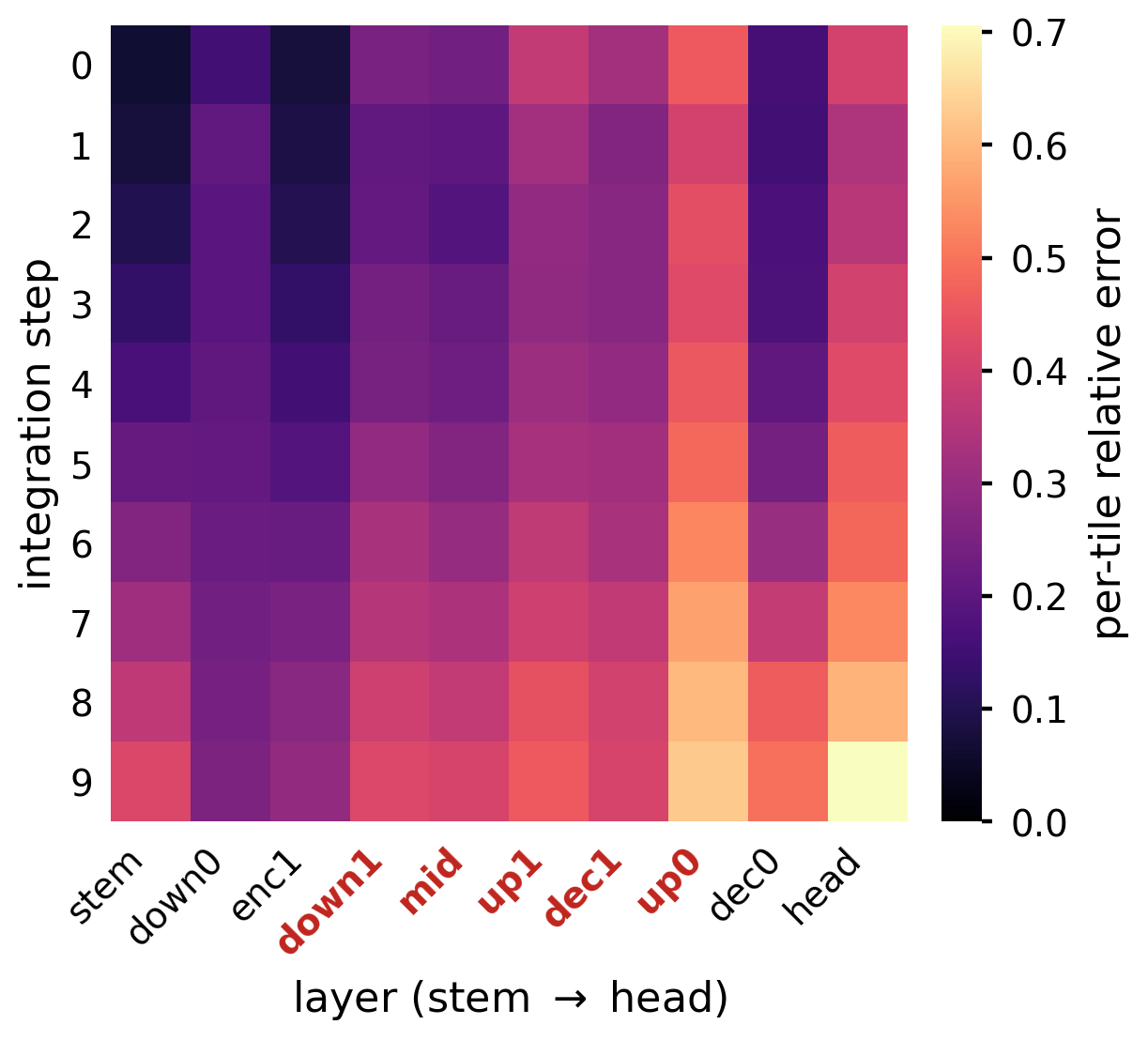}
  \caption{\textbf{The chip spends no cycles on precision the output does not need.} Per-tile error of the concurrent depth-collapse against the floating-point reference across the unrolled (step$\times$layer) graph, at the window where the generated image first reaches pixel-RMSE $<0.1$.
  }
  \vspace{-1em}
  \label{fig:cifar_depth}
\end{figure}

This points to a property of the sequential parallelism and depth-collapse worth stating in its own right, and to why the sequential baseline is so costly. The concurrent run does not spend cycles making the intermediate layers accurate. At the moment the image first becomes good (pixel-RMSE below 0.1, after only ${\approx}3$M of the eventual cycles), the layers in the middle of the network still carry 20--70\% per-tile error against the floating-point reference, and they remain inaccurate even at full convergence (\cref{fig:cifar_depth}). The output tolerates this because the integration is residual: the image is $x_0 + \sum_k \mathrm{d}t\, v_k$, so per-step, per-layer errors average out and the residual dominates. Sequential parallelism thus allocates precision automatically to where it reaches the output and spends none where it does not, which is precisely the accuracy the layer-by-layer baseline wastefully demands of every intermediate tile.

\subsection{Free-energy estimation on alanine dipeptide}
\label{sec:aldp}

Generative modelling has increasingly turned to the molecular sciences, where diffusion and flow-matching models generate small molecules, protein structures and materials~\cite{noe2019}. One problem in this space which has recently gained more interest is that of estimating free-energy difference~\citep{iclr2026_escorted}, a thermodynamic quantity. We take this as our final demonstration, on alanine dipeptide, the standard small-molecule benchmark, and find that a thermodynamic substrate is especially well matched to this thermodynamic problem.

\begin{figure}[t]
  \centering
  \includegraphics[width=0.95\columnwidth]{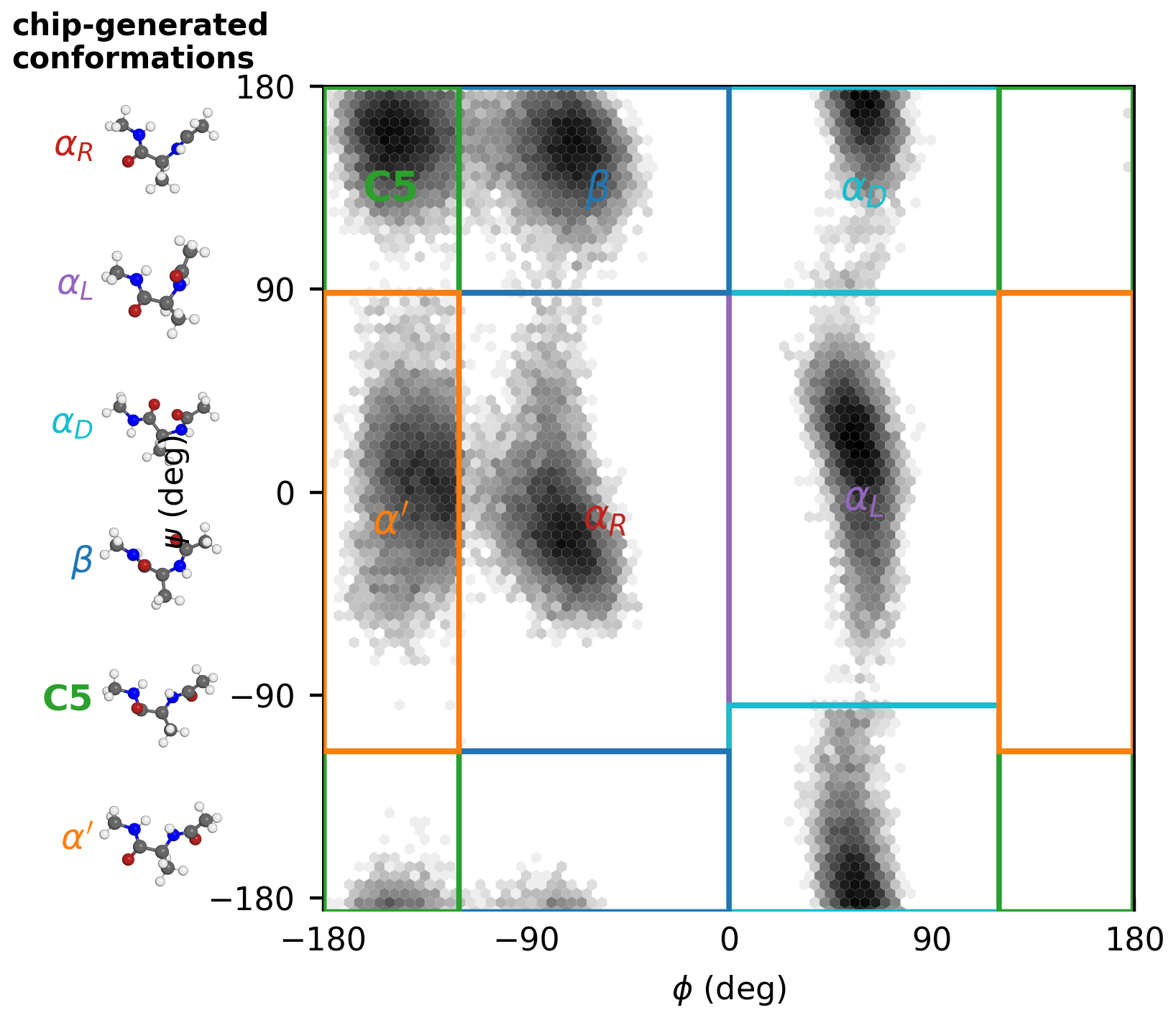}
  \caption{\textbf{Chip-generated conformations and the metastable basins.} \emph{Left:} chip-generated alanine-dipeptide conformations, one per metastable basin. \emph{Right:} density of the reference molecular-dynamics ensemble in the backbone-dihedral plane $(\phi,\psi)$, with the six flat-bottom state-definition boxes that delimit the basins overlaid.}
  \label{fig:aldp_hero}
  \vspace{-1em}
\end{figure}

The slow degrees of freedom of alanine dipeptide are the two backbone dihedrals $(\phi,\psi)$, whose Boltzmann distribution $\pi(x)\propto\exp(-U(x)/k_B T)$ concentrates in a small number of metastable basins (\cref{fig:aldp_hero}). The quantity of interest is the free-energy difference between two basins $i$ and $j$,
\begin{equation}
  \Delta F_{ij} = F_j - F_i = -k_B T \log \frac{Z_j}{Z_i},
  \label{eq:dF}
\end{equation}
which fixes their relative equilibrium population through $p_j/p_i = \exp(-\Delta F_{ij}/k_B T)$. Free-energy differences of this kind set binding affinities, solubilities and conformational preferences, and their estimation is among the most demanding tasks in computational chemistry, because the basins are separated by barriers that equilibrium molecular dynamics crosses only rarely and so converges slowly. We stress that the task here is free-energy estimation rather than conformational sampling.

\paragraph{Estimating free-energy differences.}
We estimate $\Delta F$ by targeted free-energy perturbation~\cite{jarzynski1997, jarzynski2002}, using a trained Riemannian flow-matching model~\cite{lipman2023, chenLipman2024} as the targeting map $\mathcal{T}$ that transports configurations from one basin to another. The work of transporting a configuration $x_0$ to $x_1 = \mathcal{T}(x_0)$ between basins with potentials $U_i$ and $U_j$ is
\begin{equation}
  W(x_0) = U_j\!\big(\mathcal{T}(x_0)\big) - U_i(x_0)
           - k_B T \log\big|\det \nabla \mathcal{T}(x_0)\big|,
  \label{eq:work}
\end{equation}
where the log-determinant of the map is the time integral of the divergence of the velocity field $v_\theta$ along the trajectory~\cite{chen2018node},
\begin{equation}
  \log\big|\det \nabla \mathcal{T}(x_0)\big|
    = \int_0^1 \big(\nabla\!\cdot v_\theta\big)(x_t, t)\,\mathrm{d}t.
  \label{eq:logdet}
\end{equation}
We train one such map between $\alpha_R$ and each of the other five basins, and estimate the free-energy difference of each basin to $\alpha_R$ with the Bennett acceptance ratio~\cite{bennett1976}, applied to the forward and reverse works of that single pair (\cref{app:aldp}). The remaining pairwise differences follow by subtraction, $\Delta F_{ij} = \hat F_j - \hat F_i$.

\begin{figure*}[t!]
  \centering
  \includegraphics[width=0.95\linewidth]{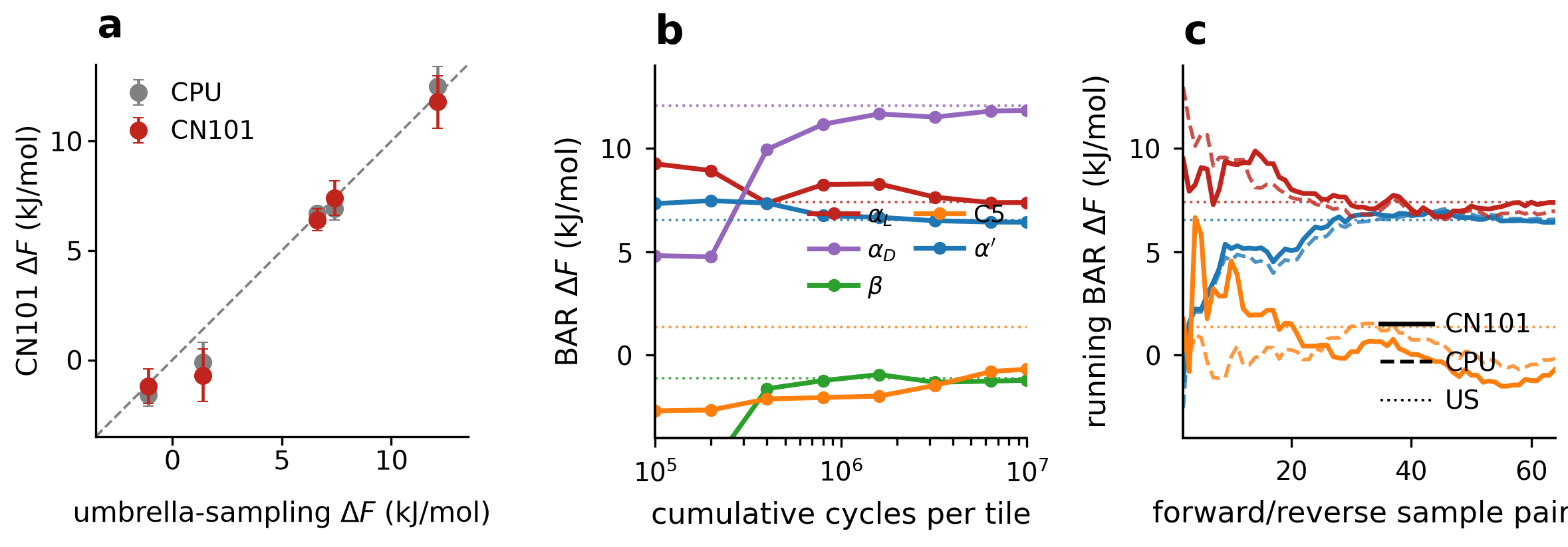}
  \caption{\textbf{Free-energy estimation on alanine dipeptide.} \textbf{(a)}~On-chip free energies for each basin relative to $\alpha_R$ against the umbrella-sampling reference (kJ/mol), with the CPU estimate of the same flows overlaid.
  \textbf{(b)}~The free-energy estimate reconstructed at each cumulative cycle budget per tile, one curve per basin.
  \textbf{(c)}~The running estimate over the chip works (solid) tracks the running estimate over the CPU works (dashed) as samples accumulate, for a tight basin ($\alpha'$), a hard one (C5) and an intermediate one ($\alpha_L$).}
  \label{fig:aldp}
\end{figure*}

\paragraph{Why sequential parallelism helps.}
The cost of this estimator is set by the log-determinant~\cref{eq:logdet} rather than by the transport~\cite{grathwohl2019}. A configuration is carried faithfully in relatively few integration steps, but the divergence integral converges only as the number of steps grows, so it is the density estimation, not the sampling, that sets the depth of the computation. On CN101, using the sequential parallelism property, the divergence trace part of the computation can equilibrate at the same time as the transport rather than in a separate pass.

\subsubsection{Setup}
We estimate $\Delta F$ between all six metastable states of Alanine Dipeptide ($\alpha_R$, $\alpha_L$, $\alpha_D$, $\beta$, C5 and $\alpha'$) through a star of flows centred on $\alpha_R$~\cite{iclr2026_escorted}, so that one reference state anchors every estimate and the remaining pairwise differences follow by subtraction. For each basin we draw $64$ equilibrium configurations from a reference molecular-dynamics trajectory, transport them to $\alpha_R$ and back in both directions using the learned flows, and combine the resulting forward and reverse works pairwise with the Bennett acceptance ratio. Each transport integrates the flow with a single Hutchinson probe and runs across six CN101 chips, its per-step matmul read out as a time-average over a schedule rising to $10^7$ cycles per tile. We compare against two references. The first is the umbrella-sampling estimate of the same free energies~\cite{torrie1977, iclr2026_escorted}, an independent physical ground truth. The second is the CPU estimate of the same flows, which isolates the contribution of the substrate from that of the model. The flows, the reference data and the training procedure are described in \cref{app:aldp}.

\subsubsection{Results}
Across the basins the chip reproduces the CPU estimate of the same flows to within $0.8$~kJ/mol (\cref{tab:aldp}, \cref{fig:aldp}a), and, with C5 a noticeable exception, the free-energy differences track the umbrella-sampling reference. Where the flow is accurate the chip is accurate; the on-chip and CPU estimates agree closely. The transported configurations are themselves valid molecular structures, as illustrated in \cref{fig:aldp_hero}.

\begin{table}[t]
  \centering
  \small
  \caption{On-chip free energies $\Delta F$ (kJ/mol, relative to $\alpha_R$) against the CPU estimate of the same flows and the umbrella-sampling reference. Uncertainties on CN101 and CPU are bootstrap standard deviations over the $64$ configurations drawn from each basin.
  }
  \vspace{-0.2cm}
  \label{tab:aldp}
  \begin{tabularx}{\linewidth}{@{}lXXX@{}}
    \toprule
    \textbf{Basin} & \textbf{CN101} & \textbf{CPU} & \textbf{US ref.} \\
    \midrule
    $\alpha_L$ & $7.4 \pm 0.8$  & $6.9 \pm 0.5$  & $7.4$ \\
    $\alpha_D$ & $11.8 \pm 1.2$ & $12.5 \pm 0.9$ & $12.1$ \\
    $\beta$    & $-1.2 \pm 0.8$ & $-1.6 \pm 0.5$ & $-1.1$ \\
    C5         & $-0.7 \pm 1.2$ & $-0.1 \pm 0.9$ & $1.4$ \\
    $\alpha'$  & $6.4 \pm 0.5$  & $6.7 \pm 0.2$  & $6.6$ \\
    \bottomrule
  \end{tabularx}
\end{table}

\Cref{fig:aldp}b shows how CN101 carries each value as a time-average, such that the free energy converges with the cycle budget rather than resolving at once. As the schedule rises to $10^7$ cycles per tile the substrate noise averages out and the estimate settles onto its converged value. However, the free energy is an exponential average of the per-sample works~\cite{zwanzig1954}, so it does not simply average the substrate noise away as an ordinary mean would. Instead, residual noise can bias the estimate. Here that bias is however small and shrinks as the cycle budget grows, and combining forward and reverse transports suppresses it further, so the chip and CPU running estimates track each other as samples accumulate (\cref{fig:aldp}c).


\section*{Discussion}

In this work we have given a substrate-independent formalisation of equilibration-style thermodynamic computing, and a digital instantiation of it in silicon as the CN101 prototype chip. Across four workloads, spanning linear systems, conditional MNIST generation, a convolutional flow on CIFAR-10, and molecular free-energy estimation, we have shown how the time-averaged readout $\yhatT$ converges to the intended function as the window grows, which is the anytime-precision axis realised in hardware. We characterise the other two axes directly: parallel sample aggregation on the VAE decoder, and sequential parallelism on both the within-chip VAE decoder and the cross-chip flow-matching pipeline.

A recurring feature of the experiments is a small bias floor in the relative error of $\yhatT$ at large $T$, where the variance has been averaged away and a residual offset remains. On the linear-algebra workload this floor sits near one percent across the dimensions tested, and on the generative models it inherits a further contribution from the per-layer encoding error that accumulates through the network. An analogous floor was reported on the analogue Stochastic Processing Unit and attributed there to circuit non-idealities~\cite{melanson2025}. On CN101 the dominant contribution is instead the finite precision of the weight registers. Distinct from this equilibrium floor is a transient, or burn-in, bias. Unlike the variance, this transient is shared across independent replicas started from the same state, so parallel sample aggregation does not remove it. Both the equilibrium floor and this transient are properties of this instantiation rather than of the formulation.

The principal open question these results raise is how to design generators that both equilibrate and mix quickly, with short $\trelax$ and $\tcorr$~\cite{levinPeresWilmer2017}, on a digital substrate. One route is suggested by the field whose name the paradigm borrows. Rather than wait for a generator to equilibrate, the tools of stochastic and non-equilibrium thermodynamics drive it through a finite-time protocol and recover the equilibrium answer by reweighting the trajectories with the fluctuation theorems of Jarzynski and Crooks~\cite{jarzynski1997,crooks1999,seifert2012}. Finite-time thermodynamics bounds the dissipation such a protocol incurs through the geometry of optimal driving~\cite{sivak2012, crooks2007}, and recent work shows that the protocols can be learned~\cite{iclr2026_escorted}. A learned non-equilibrium protocol is, in our opinion, the most promising route past the mixing-time barrier.

CN101 is a prototype, and the first chip in a broader architectural programme aimed at production-scale generative AI. The present paper characterises the behaviour of the formulation on this first generation, while the energy and latency targets that motivate the programme are deferred to later chips, on which the blocks left deliberately first-generation here are to be improved. Two choices in CN101 already anticipate that path. Its modular pseudo-random-number interface accepts a physical noise source in a future chip without a change to the datapath, and its polysynchronous clocking removes the global clock-distribution constraint that would otherwise bound how far a single chip, or an assembly of them, can scale. That a standard generative model, trained and run on GPUs and mapped onto the chip without custom layers, runs correctly on dynamics this far removed from a Langevin system is the result we find most telling: it indicates that thermodynamic computing in its equilibration-style formulation, and the hardware advantages that come with it, are within reach of the models the field already builds.

\section*{Acknowledgements}
Normal Computing thanks the Advanced Research and Invention Agency’s (ARIA) Scaling Compute programme for funding this work.

\bibliography{bibliography}
%

\appendix
\crefalias{section}{appendix}
\crefalias{subsection}{subappendix}

\begin{center}
  {\normalfont\bfseries\fontsize{14}{17}\selectfont Appendix}
\end{center}


\section{Formal statement of the equilibration formalism}\label{app:formal}

This appendix provides formal statements of the claims made in \cref{sec:framework}. We restate the formalism under formal assumptions and show that consistency is preserved under both parallel aggregation and sequential composition.

\paragraph{Formalism and ergodicity.}
Let $S$ be a measurable state space, and let $(S_t)_{t \geq 0}$ be a discrete-time Markov chain on $S$ with transition kernel $P_x$ that depends parametrically on the input $x$ to the computation. The dynamical generator $\Lstar$ is the forward operator associated with $P_x$, acting on distributions by $\Lstar\mu = \mu P_x - \mu$. Suppose $P_x$ admits a unique stationary distribution $\pi_x$ satisfying $\Lstar \pi_x = 0$, equivalently $\pi_x P_x = \pi_x$, and suppose the chain is ergodic with respect to $\pi_x$. Let $\varphi : S \to \RR^m$ be a measurable observable with $\Epi[\|\varphi(S)\|] < \infty$. The computation is
\begin{equation*}
  f(x) \;\triangleq\; \Epi[\varphi(S)],
\end{equation*}
and the chip's output for a run of length $T$ is
\begin{equation*}
  \hat{y}_T \;=\; \frac{1}{T} \sum_{t=1}^{T} \varphi(S_t).
\end{equation*}
By the ergodic theorem~\cite{meyn2009markov}, $\hat{y}_T \to f(x)$ almost surely as $T \to \infty$, regardless of the distribution of $S_0$. As is standard for Markov chain averages, the variance of $\hat{y}_T$ falls as $\tcorr/T$ with the integrated correlation time $\tcorr$, while before stationarity the estimator carries a burn-in bias that decays once $T$ exceeds the relaxation time $\trelax$~\cite{levinPeresWilmer2017}.

\paragraph{Parallel aggregation.}
Consider $K$ independent runs of the same chain, each yielding a per-run estimator $\hat{y}_T^{(k)}$ as above. The aggregate estimator is
\begin{equation*}
  \bar{y}_T^{(K)} \;=\; \frac{1}{K} \sum_{k=1}^{K} \hat{y}_T^{(k)},
\end{equation*}
and independence of the runs gives $\EE[\bar{y}_T^{(K)}] = \EE[\hat{y}_T^{(1)}]$ and $\mathrm{Var}(\bar{y}_T^{(K)}) = \mathrm{Var}(\hat{y}_T^{(1)}) / K$, so the rate of variance reduction is set by the aggregate sample throughput $KT$ and the runs need no coordination beyond the final average.

\paragraph{Sequential parallelism.}
The framework permits the dynamics to be split across subsystems that share no clock and exchange only aggregated outputs. Consider an $L$-stage pipeline in which stage $\ell$ implements a thermodynamic computation $f_\ell(x_\ell) = \EE_{\pi_\ell}[\varphi_\ell(S^{(\ell)}) \mid x_\ell]$, with $x_{\ell+1}$ produced from the output of stage $\ell$. Let $\hat{y}_T^{(\ell)}$ denote stage $\ell$'s time-averaged readout, run on the upstream estimate $\hat{y}_T^{(\ell-1)}$, and assume each stage is consistent in the sense that $\hat{y}_T^{(\ell)} \to f_\ell(x_\ell)$ almost surely whenever its input converges to $x_\ell$. If each $f_\ell$ is continuous in its argument, applying this inductively along the pipeline gives
\begin{equation*}
  \hat{y}_T^{(L)} \;\longrightarrow\; (f_L \circ \cdots \circ f_1)(x_1) \quad \text{a.s.}
\end{equation*}
as the per-stage windows grow, by the extended continuous-mapping theorem~\cite{vanderVaart1998}. Consistency is therefore propagated through downstream stages, and the argument requires only stage-level consistency and continuity of the per-stage maps, not a shared clock between stages.


\section{Compiling a neural network on CN101}\label{app:mapping}

Every workload is run on CN101 by the same recipe.
The network is trained off-chip in floating point, without custom layers, and its weights are quantised only at deployment, to $8$-bit integers in $64\times64$ blocks loaded into the stochastic matrix multipliers.
A layer that fits a single tile occupies one block; a wider or deeper layer is partitioned into $64\times64$ blocks and assigned across tiles and chips, the partition being the only model-dependent choice.

When a model fits the four tiles of a single chip, as the VAE decoder does, each neuron bank applies its layer's bias and ReLU on-chip and the cascade runs as one dispatch; when a model is spread across tiles and chips, the chip evaluates the $64\times64$ matmuls and the host applies the bias, normalisation, activation and any integration step between windows.


\subsection{Composition and multi-chip execution}\label{app:multichip}

Every concurrent configuration in the paper exercises the sequential parallelism of \cref{sec:scaling}: each stage advances on the still-converging estimate of the stage feeding it rather than waiting for it to settle.

Within a chip this exchange is continuous. The four tiles run on independent ring-oscillator clocks, with cross-tile signals synchronised at bank boundaries, and exchange bitstreams over the stochastic streaming network. A four-layer cascade of tiles therefore runs as a single-chip dispatch with one final readout and no host intervention between layers.

Across chips the same composition is realised window by window. Over a window of $\Delta$ cycles each chip advances on its current input; the host then reads each chip's running estimate from its output accumulator, applies any host-side bias, normalisation or integration step, and refreshes the downstream inputs before the next window. The sync interval $\Delta$ can be chosen between fully sequential execution and the fully concurrent regime $\Delta=1$. Integrated multi-chip hardware, on which the estimate exchange is routed on-board rather than passed through the host, is left to a subsequent chip in the Carnot Architecture.


\section{Conditional VAE: training and deployment}\label{app:vae}

The decoder architecture, dataset and tile mapping are given in \cref{sec:vae}; here we record the training and deployment detail. The encoder, used only during training, mirrors the decoder and outputs the mean and log-variance of the $16$-dimensional latent. The network is trained end to end in floating point with the reparametrisation trick and a class-conditional evidence lower bound~\cite{sohn2015cvae}, the reconstruction and Kullback--Leibler terms weighted equally ($\beta=1$). Training uses the Adam optimiser at learning rate $10^{-3}$ and batch size $128$ for $100$ epochs; no CN101-specific regularisation is applied, and the weights are quantised to $8$-bit integers only at deployment.

For inference the weights are held stationary for the duration of a run. The per-$T$ operating points of \cref{fig:vae}c,d are selected by a search over the chip's two configurable knobs that do not change the trained model: the input scale applied to each tile and the weight-register precision. For each cycle budget $T$ the configuration minimising the scale-invariant RMSE against the reference is chosen, so that the reported curve traces the best operating point available at each $T$ rather than a single fixed setting.


\section{Convolutional flow on CIFAR-10: training and schedule}\label{app:cifar}

The \textsc{ChipUNetTiny} architecture, the shift-then-$1\times1$ convolution mapping and the tile pinning are given in \cref{sec:cifar}; here we record the training and readout schedule.
The velocity field is trained by flow matching~\cite{lipman2023} on ten CIFAR-10 images with the Adam optimiser at learning rate $2\times10^{-4}$ and full-batch gradients for $20{,}000$ steps, and is then reflowed once~\cite{liu2022rectified}: samples drawn from the trained model are paired with their originating noise and the field is retrained on the straightened couplings for a further $20{,}000$ steps, so that generation needs only $N=10$ integration steps.
The model is deliberately overfit to these ten images, the prototype's on-chip weight memory being too small to hold a general CIFAR generator, so the experiment establishes that the substrate runs the deep convolutional flow correctly rather than that this network generalises.

At inference every per-step matmul is read out as a time-average over a schedule rising to $5\times10^{5}$ cycles per tile, with the first windows, over which the unrolled relaxation fills, discarded as a burn-in before averaging.


\section{Alanine dipeptide: model, data, and free-energy protocol}\label{app:aldp}

\paragraph{Reference data.}
We model alanine dipeptide in implicit solvent and describe each configuration in internal bond--angle--torsion coordinates~\cite{noe2019}: $21$ bonds and $20$ angles on $\mathbb{R}^{41}$, and $19$ backbone and side-chain dihedrals on the torus $\mathbb{T}^{19}$. The six metastable basins ($\alpha_R$, $\alpha_L$, $\alpha_D$, $\beta$, C5 and $\alpha'$) are defined by flat-bottom boxes in the backbone-dihedral plane $(\phi,\psi)$ (\cref{fig:aldp_hero}), following the state definitions of \citet{iclr2026_escorted}. For each basin we draw equilibrium configurations from a restrained molecular-dynamics trajectory ($20$\,ns per basin, a flat-bottom restraint of strength $k=100$\,kJ\,mol$^{-1}$\,rad$^{-2}$ confining the dynamics to the basin). These restrained ensembles are the reference distributions the flow transports between, and the unrestrained potential $U(x)$ supplies the energies entering the work.

\paragraph{Flow-matching model.}
For each pair $(\alpha_R, X)$ we train a Riemannian flow-matching model~\cite{lipman2023, chenLipman2024} on the product geometry $\mathbb{T}^{19}\times \mathbb{R}^{41}$, with the dihedral block carried through a $\cos/\sin$ embedding so that the velocity field respects the periodicity of the torus. The velocity field is a residual multilayer perceptron of two blocks of width $128$ with layer normalisation and ReLU activations, conditioned on the integration time. It is trained to transport the restrained $\alpha_R$ ensemble onto the restrained $X$ ensemble, with the reverse map the same flow integrated backwards, using the Adam optimiser at learning rate $10^{-3}$ and batch size $256$ for $200$ epochs.

\paragraph{Free-energy estimation.}
We estimate free-energy differences by targeted free-energy perturbation~\cite{jarzynski2002}, with the trained flow as the targeting map $\mathcal{T}$ and the work of a transport given by~\cref{eq:work} at temperature $T = 300$\,K, its log-determinant estimated with a single Hutchinson probe per integration step~\cite{hutchinson1989} (rank one). We train a star of five maps, one between the reference basin $\alpha_R$ and each of the other basins, following the construction of \citet{iclr2026_escorted}. For each pair $(\alpha_R, X)$ we estimate the free-energy difference $\Delta F(\alpha_R, X)$ with the Bennett acceptance ratio~\cite{bennett1976}, which combines the forward works of the transport $\alpha_R\!\to\!X$ with the reverse works of $X\!\to\!\alpha_R$. With $\alpha_R$ as the common reference this fixes each free energy $\hat F_i$ relative to it, and the remaining pairwise differences follow by subtraction, $\Delta F_{ij}=\hat F_j-\hat F_i$. Each free-energy difference draws $64$ configurations from each basin, and uncertainties are bootstrap standard deviations over those samples.

\paragraph{Chip deployment.}
The flow is deployed on the CN101 cascade with its weights quantised in $64\times64$ blocks. Each transport integrates $32$ Euler steps, with the Hutchinson tangent carried in the same forward sweep rather than a separate pass, and every per-step matmul is read out as a time-average over a twenty-window schedule whose per-window length rises geometrically to $10^{7}$ cycles per tile (\cref{fig:aldp}b).


\section{Reference computation, metrics and reproducibility}\label{app:repro}
Each chip result is compared against a reference that runs the same $8$-bit-quantised model in exact floating-point arithmetic on the host. Because reference and chip share the model and differ only in the substrate, the comparison isolates the contribution of the substrate from that of the model.

Errors are reported per workload in the natural metric. For the VAE the output is scored by a scale-invariant RMSE against the reference, computed after a single global scale and offset are removed, so that the metric reflects structural rather than overall-gain error. For the convolutional flow the metric is the pixel-RMSE of the generated image against the reference. For alanine dipeptide it is the free-energy difference recovered through the Bennett acceptance ratio.

The randomness each run requires is drawn from the on-chip generators with seeds verified to produce non-overlapping sequences across neurons and across runs, and the independent replicas combined in parallel sample aggregation use distinct such seeds. Error bars are stated with each result: the VAE figures average over ten digit classes and ten latent draws per class, and the free-energy estimates use bootstrap standard deviations over the drawn configurations.

\end{document}